\documentclass[12pt]{iopart}
\usepackage{graphics}
\usepackage{epsfig}
\usepackage{amsfonts}
\usepackage{color}
\usepackage{psfrag}

\begin{document}
\title[]{Lightlike Geodesics and Gravitational Lensing in the Spacetime of an Accelerating Black Hole}
\author{Torben C Frost and Volker Perlick}
\address{ZARM, University of Bremen, 28359 Bremen, Germany.
\\
Email:
torben.frost@zarm.uni-bremen.de, volker.perlick@zarm.uni-bremen.de}
\vspace{10pt}
\begin{indented}
\item[]January 20, 2021
\end{indented}

\begin{abstract}
The C-metric is a solution to Einstein's vacuum field equation
that describes an accelerating black hole. In this paper we discuss
the propagation of light rays and the resulting lensing features
in this metric. We first solve the lightlike geodesic equation using
elliptic integrals and Jacobi elliptic functions. Then we fix a
static observer in the region of outer communication of the C-metric and introduce
an orthonormal tetrad to parameterise the directions of the light rays ending at
the position of the observer using latitude-longitude coordinates on the observer's
celestial sphere. In this parameterisation we rederive the angular radius of the
shadow, we formulate a lens equation, and we derive the redshift and the travel
time of light rays. We discuss the relevance of our theoretical results for
detecting accelerating black holes described by the C-metric and for
distinguishing them from non-accelerating black holes.

\end{abstract}
\maketitle
\section{Introduction}
In 2019 the Event Horizon Telescope (EHT) Collaboration released images showing
the shadow of the supermassive black hole in the core of the galaxy M87 observed
in April 2017 \cite{EHTCollaboration2019}. Besides being a significant scientific
achievement in itself this observation also provided an additional confirmation
of general relativity: The observed features were found to be in agreement with
the assumption that at the centre of M87 there is a Kerr black hole.

Currently the resolution of the EHT is limited by the distribution of radio telescopes
on Earth and thus extension of Very Long Baseline Interferometry (VLBI) to space
is needed for further increasing its resolution. Space VLBI has a tradition since
the late 1970s and made particular progress with the Spektr-R satellite
that was part of the RadioAstron programme \cite{Kardashev2013,Kardashev2017}
and terminated operation in 2019. However, for the EHT one needs radio
telescopes that can operate at millimeter or submillimeter wavelengths.
Space VLBI with telescopes in this range has not been done so far, but projects
that are attempting to do this are now at the planning stage. When realised,
this will allow us observing the shadow and other lensing features of supermassive
black holes with unprecedented accuracy. In particular, it will then be possible to use
this kind of observations for discriminating between different black hole models.
On the basis of general relativity, the Kerr metric or, in the case of vanishing
rotation, the Schwarzschild metric is the standard model for an uncharged
black hole. However, Einstein's vacuum field equation admits more general
black hole solutions. Their astrophysical relevance is unclear because they
are exotic in the sense that the spacetime outside of the horizon is either
not free of singularities or not asymptotically flat (or both). Future
observations could make it possible to distinguish Kerr black holes
from these exotic black holes by their lensing features. For this  reason,
it is of relevance to theoretically study the lensing features of exotic black
holes.  Moreover, such investigations are of some interest by itself because
they illustrate what kind of mathematical peculiarities are allowed by the
vacuum field equation.

In this paper we want to discuss the lensing features of one of these
exotic solutions which is known as the C-metric. It was found already
in 1918 by Levi-Civita \cite{LeviCivita1918} and belongs to the family
of Pleba{\'n}ski-Demia{\'n}ski electrovacuum spacetimes with cosmological constant
\cite{Plebanski1976}. The name C-metric refers to a review by Ehlers and Kundt
\cite{Ehlers1962} who rediscovered this metric within three classes of vacuum
solutions they labelled A, B and C. The C-metric is discussed in detail in an
article by Griffiths et al. \cite{Griffiths2006} and in a book by Griffiths and
Podolsk{\' y} \cite{Griffiths2009}. As outlined in these works, the C-metric
describes an accelerating black hole. (The  maximal analytical extension actually
describes two black holes accelerating in opposite directions; however,
as these two black holes live in causally disconnected regions, it is
reasonable to restrict the consideration to one of them.) The metric depends
on two parameters, a mass parameter $m$ and an acceleration parameter
$\alpha$. For $\alpha =0$ it reduces to the Schwarzschild metric.
This is in agreement with the fact that for a black hole in
uniform motion the metric of the ambient spacetime can be reduced
to the Schwarzschild metric by a Lorentz transformation. Note,
however, that this Lorentz transformation becomes singular if
the velocity of the black hole approaches the speed of light,
see Aichelburg and Sexl \cite{Aichelburg1971}.

In the C-metric the acceleration of the black hole is caused either by a string
that pulls the black hole or by a strut that pushes it (or by both). The string
and the strut manifest themselves as conical singularities; one of them, but not
both, could be removed by changing the periodicity of the $\varphi$ coordinate,
as was first observed by Kinnersley and Walker\cite{Kinnersley1970}, cf.
Griffiths and Podolsk{\' y} \cite{Griffiths2009}. Recently
Kofro{\v n} \cite{Kofron2020} interpreted the string and the strut as a
null dust. He associated the asymmetry of the C-metric with a momentum
flux towards the black hole through the strut and away from the black
hole through the string. While one can hardly expect a black hole with such
a string or a strut attached to be literally realised in Nature,  there might be
accelerating black holes that can be described by such a model to within a
reasonable approximation. In any case, history has taught us that solutions
to Einstein's vacuum field equation should be taken seriously.

Whereas the topological and causal structure of the C-metric has been
investigated by many authors, the geodesics in this metric have been
studied only partially. Farhoosh and Zimmerman \cite{Farhoosh1980}
investigated the motion on radial timelike geodesics. In 2001 Pravda and Pravdova
\cite{Pravda2001} analytically derived relations for describing the circular motion
of massive and massless particles (photons) about the axis. In the same year
Chamblin \cite{Chamblin2001} qualitatively extended their analysis to circular
motion in the C-metric with a negative cosmological constant. Both studies came
to the conclusion that circular null geodesics are unstable. Podolsk{\' y} et al.
\cite{Podolsky2003} discussed several special features of the null geodesic equation
in the C-metric with a negative cosmological constant; however, they provided an
explicit exact solution only for geodesics with, in their notation, $J=Q=0$. Bini et al.
\cite{Bini2005} investigated the motion of spinning particles on circular geodesics
about the axis with the C-metric as background. The most detailed study of timelike
and null geodesics in the C-metric was provided by Lim \cite{Lim2014} who used a
combination of analytical and numerical methods. However, his study mainly focused
on timelike geodesics and provided explicit analytic solutions only for the radius
coordinate on the axes and the time coordinate of radial null geodesics.

Of particular interest in the C-metric is the photon sphere,
which is filled with unstable null geodesics. Its existence gives rise to the shadow
of a black hole that can be observed by VLBI. Grenzebach et
al. \cite{Grenzebach2015a,Grenzebach2016} determined in Boyer-Lindquist-like
coordinates the shadow of an accelerating black hole with additional spin and NUT
parameter. Specifying their equations (22), (27a) and (29)
to the case of vanishing spin and vanishing NUT parameter gives the photon
sphere and the angular radius of the shadow in the C-metric. In Hong-Teo
coordinates \cite{Hong2003} the photon sphere in the C-metric was investigated
by Gibbons and Warnick \cite{Gibbons2016}. In particular, they showed that its
intrinsic geometry is not that of a ``round sphere'' (i.e., of a sphere isometrically
embedded into Euclidean 3-space) but that it has at least one conical singularity.
They also plotted parts of a geodesic on that surface; however, they did not
explicitly discuss analytic solutions to the equations of motion. Recently Alrais
Alawadi et al. \cite{AlraisAlawadi2021} rederived, in Boyer-Lindquist-like
coordinates, the coordinate radius of the photon sphere and they also
determined the coordinate angle of what we will refer to as the photon cone.
In addition they performed a stability analysis and found that the circular null
geodesic at the intersection of the photon sphere and the photon cone is unstable
with respect to radial perturbations. Actually, all lightlike geodesics in the photon
sphere are unstable with respect to radial perturbations as will be discussed
below.

So it is fair to say that a comprehensive presentation of analytical solutions to
the lightlike geodesic equation in the C-metric is not yet available in the literature.
It is one of the main goals of this paper to provide such a presentation.
Such analytical solutions have the benefit of clearly showing the influence of each
parameter onto the geodesics. This is a major advantage over numerical solutions.
Moreover, it was argued by several authors, e.g. by Yang and Wang \cite{Yang2013}
who considered light rays in the Kerr spacetime, that an analytical representation
of the geodesics is useful also for ray-tracing codes because it reduces the
computation time.

Therefore the first objective of this paper is to analytically solve the equations of
motion for lightlike geodesics in the C-metric. We approach this problem by using
Jacobi's elliptic functions and Legendre's elliptic integrals. For the Schwarzschild
spacetime, this approach has a very long tradition, beginning in 1920 with a
short paper by Forsyth \cite{Forsyth1920} and a detailed investigation by
Morton \cite{Morton1921}. Jacobi's elliptic functions were also used in 1959
by Darwin \cite{Darwin1959} who discussed important features of lightlike and
timelike geodesics in the Schwarzschild metric and, much more recently, for
lightlike geodesics in the Kerr metric by Yang and Wang \cite{Yang2013} and by
Gralla and Lupsasca \cite{Gralla2020}.

The analytical solutions of the lightlike geodesic equation in the C-metric will
allow us to study, in the second part of the paper, the relevant lensing features
of an accelerating black hole. We will calculate the angular radius of the shadow,
we will set up a lens equation, and we will discuss the redshift and the travel time
of light in the C-metric. Up to now there are only partial results in this direction
available in the literature. We have already mentioned that Grenzebach et al.
\cite{Grenzebach2015a,Grenzebach2016} discussed the shadow in a class of spacetimes
that contains the C-metric as a special case. Sharif and Iftikhar \cite{Sharif2016}
considered the bending angle of light rays in the equatorial plane of the C-metric
with additional spin and NUT parameter. This analysis does not apply to the case
of the pure C-metric (without spin and without NUT parameter) because in this case
the only lightlike geodesics in the equatorial plane are the radial ones, as
will become clear later in this paper.
Alrais Alawadi et al. \cite{AlraisAlawadi2021} calculated the bending angle of
light rays on the photon cone, however, since the angle of the photon cone is
unique for any chosen value of the acceleration parameter $\alpha$ this calculation
only considers a very limited set of null geodesics.

The remainder of the paper is structured as follows. In Section \ref{sec:Cmetric}
we summarise, for the reader's convenience, the main properties of the
C-metric. In Section \ref{sec:geo} we discuss and solve the equations of motion
for lightlike geodesics. In Section \ref{sec:lensing} we discuss the main lensing
features in the C-metric, i.e., the angular radius of the shadow, the lens equation,
the redshift and the travel time of light. In addition we address the question of
how the influence of the acceleration parameter can actually be measured.

Throughout the paper we use geometrical units such that $c=G=1$. The
convention for the metric signature is $(-,+,+,+)$.

\section{The C-metric}\label{sec:Cmetric}
The C-metric depends on two parameters, a mass parameter $m$ with
the dimension of a length and an acceleration parameter $\alpha$ with
the dimension of an inverse length. For $\alpha =0$ it reduces to the
Schwarzschild metric. In Boyer-Lindquist-like coordinates the metric
reads \cite{Griffiths2006,Griffiths2009}
\begin{eqnarray}\label{eq:Cmetric}
\fl \qquad
g_{\mu \nu}(x) dx^{\mu} dx^{\nu} = \frac{1}{\Omega (r, \vartheta )^2}
\left( - Q(r) dt^2 + \frac{dr^2}{Q(r)} + \frac{r^2d \vartheta ^2}{P(\vartheta )}
+ r^2 P( \vartheta ) \, \mathrm{sin} ^2 \vartheta \, d \varphi ^2 \right),
\end{eqnarray}
where
\begin{eqnarray}\label{eq:Omega}
\Omega (r, \vartheta ) = 1 - \alpha \, r \, \mathrm{cos} \, \vartheta \, ,
\end{eqnarray}
\begin{eqnarray}\label{eq:Q}
Q (r ) = \Big( 1 - \alpha ^2 r ^2 \Big) \left( 1 - \frac{2m}{r} \right)  ,
\end{eqnarray}
\begin{eqnarray}\label{eq:P}
P (\vartheta ) = 1 - 2 \, \alpha \, m \, \mathrm{cos} \, \vartheta \, .
\end{eqnarray}
Throughout this article, we follow the convention in \cite{Griffiths2009}. We
assume $m>0$ and $\alpha > 0$, treating the case $\alpha =0$ as a limit. Excluding
negative values of $\alpha$ is no restriction of generality because the C-metric
with $- \alpha$ is isometric to the C-metric with $\alpha$, where the isometry is
given by the reflection at the equatorial plane, $\vartheta \mapsto \pi - \vartheta$.
Note, however, that some authors, e.g. Griffiths et al. \cite{Griffiths2006},
call $\alpha$ what we call $- \alpha$.

The C-metric is static and axisymmetric and thus has two Killing vector fields
$K_{t}=\partial_{t}$ and $K_{\varphi}=\partial_{\varphi}$. The vector field $K_{t}=\partial_{t}$
is associated with the invariance of the metric under time translations and generates
a boost symmetry. The vector field $K_{\varphi}=\partial_{\varphi}$ is associated
with the invariance of the metric under rotations about the axis of symmetry
$\mathrm{sin}\, \vartheta=0$. As already mentioned in the Introduction,
the C-metric is usually interpreted as describing an accelerating black hole; the
justification for this interpretation is discussed in detail in the book by
Griffiths and Podolsk{\' y} \cite{Griffiths2009}. Observers on $t$-lines move with
the black hole in such a way that they see a time-independent metric.

The $t$ coordinate ranges over all of $\mathbb{R}$ and the angle
coordinates $\vartheta$ and $\varphi$ are assumed to have their
usual range as standard coordinates on the 2-sphere; in particular,
$\varphi$ is assumed to be $2 \pi$-periodic. Then there is a conical
singularity on the axis $\mathrm{sin} \, \vartheta =0$, i.e., the condition
of elementary flatness is violated. For $\alpha >0$ a plane that crosses
the half-axis $\vartheta = 0$ has a deficit angle and a plane that
crosses the half-axis  $\vartheta = \pi$ has a surplus angle. This may
be interpreted as saying that at $\vartheta = 0$ there is a string (with
a tension, that pulls the black hole) and that at $\vartheta = \pi$ there
is a strut (with a pressure, that pushes the black hole), see e.g. Griffiths
et al. \cite{Griffiths2006,Griffiths2009}. By giving the $\varphi$ coordinate a
periodicity different from $2 \pi$ one could remove the conical singularity
on one of the two half-axes, but not on both. Several authors, e.g.
Griffiths and Podolsk{\' y} \cite{Griffiths2009}, advocate the idea of
removing the conical singularity at $\vartheta = \pi$. We will not do
this here, so that the reader can see the effect both of the string and of
the strut. In view of the range of the angular coordinates it is also
important to note that $P(\vartheta)$ must be positive. If $2m \alpha > 1$,
this bounds the $\vartheta$ coordinate away from the axis
$\mathrm{sin} \, \vartheta=0$ which seems to be unphysical.
Therefore, we will soon restrict $\alpha$ to values such that
$2 m \alpha < 1$.

\begin{figure}[ht]
  \centering
		\includegraphics[width=0.8\textwidth]{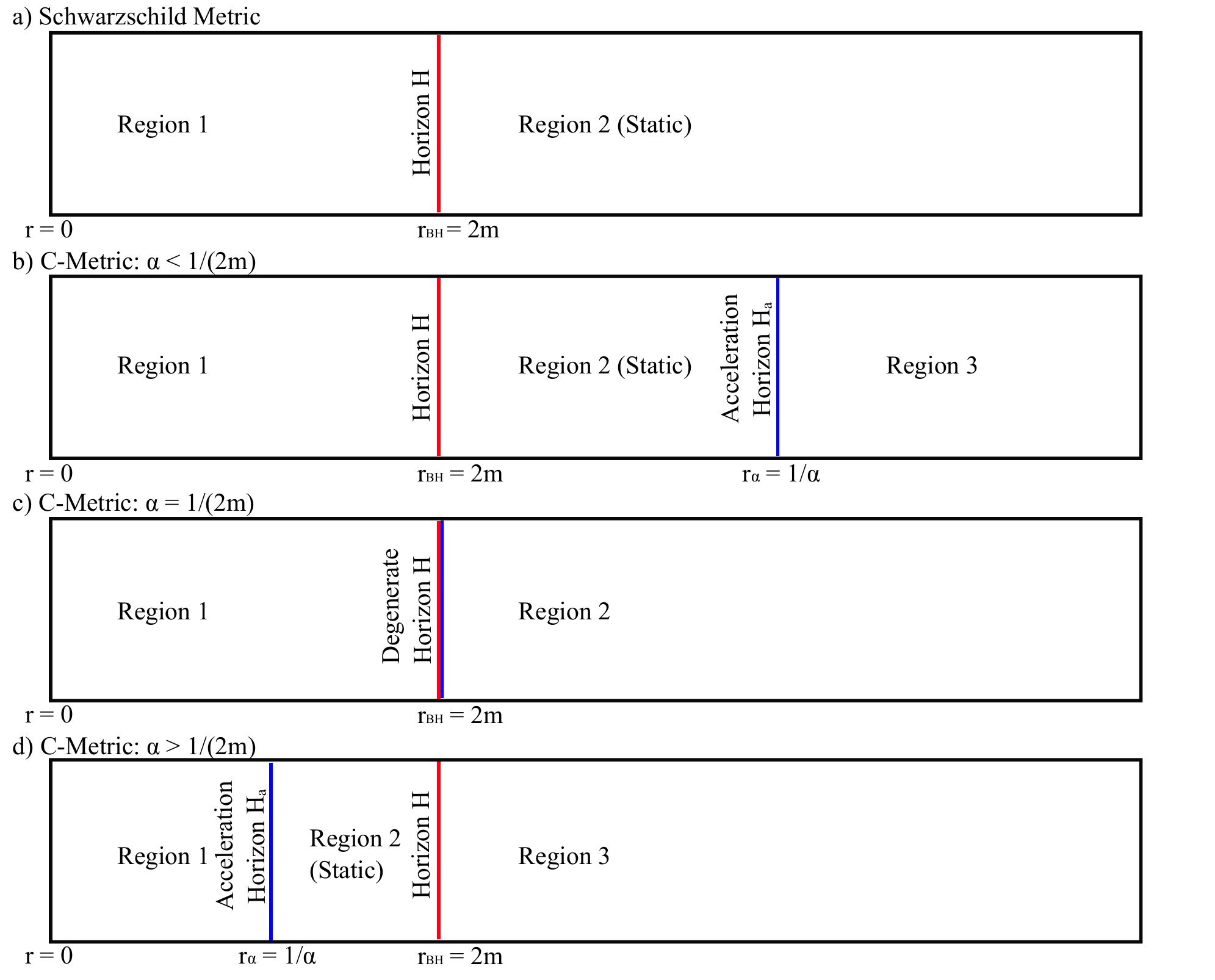}
	\caption{Position of the curvature singularity at $r=0$ and the coordinate
  singularities in a) the Schwarzschild spacetime and the C-metric for b)
  $r_{BH}<r_{\alpha}$, c) $r_{BH}=r_{\alpha}$ and d) $r_{\alpha}<r_{BH}$.
  Note that the angular coordinates are suppressed and other singularities
  are not shown.}
	\label{fig:BHHS}
\end{figure}

For discussing the range of the $r$ coordinate we have to investigate
at which $r$-values the metric, in the Boyer-Lindquist-like coordinates,
becomes singular. This is the case if $r=0$, $\Omega(r,\vartheta)=0$,
or $Q(r)=0$. Analogously to the Schwarzschild metric, the C-metric
has a curvature singularity at $r=0$, limiting the motion of particles
and light rays to the domain $r>0$. The conformal factor
$\Omega(r,\vartheta)$ vanishes when \cite{Griffiths2006}
\begin{eqnarray}
r=\frac{1}{\alpha\cos\vartheta}
\end{eqnarray}
which is possible, as $r>0$, only if $0\leq\vartheta<\frac{\pi}{2}$.
This singularity corresponds to conformal infinity. It starts at $r=1/\alpha$
for $\vartheta=0$ and stretches out to $r=\infty$ for $\vartheta=\frac{\pi}{2}$.
For $\frac{\pi}{2}\leq\vartheta\leq\pi$ the $r$ coordinate extends to infinity.
The function $Q(r)$ vanishes at $r_{BH}=2m$ and at $r_{\alpha}=
\frac{1}{\alpha}$. Both are coordinate singularities which can be removed
using appropriate coordinate transformations (see, e.g., Griffiths et
al. \cite{Griffiths2006}). These coordinate singularities lead to the
existence of horizons. Here we have to distinguish three different cases.
Figure~\ref{fig:BHHS} shows the horizon structure for the three cases
and for the Schwarzschild metric. Note that the angular coordinates are
suppressed and that, correspondingly,  the singularities at
$\Omega (r , \vartheta ) =0$, $\mathrm{sin} \, \vartheta =0$
and (possibly) $P (\vartheta ) = 0$ are not shown.

In case one (figure~\ref{fig:BHHS}b) we have $r_{BH}<r_{\alpha}$. In this case
the coordinate singularity at $r_{BH}$ marks the position of the black hole
horizon while the second coordinate singularity marks the position of an outer
horizon. Because the existence of the outer horizon arises from a non-vanishing
$\alpha$ it is usually referred to as the acceleration horizon. In the region between
the two horizons $\partial_{t}$ is timelike while $\partial_{r}$ is spacelike.
Thus this region is static. In the other two regions the causal character
of $\partial_{t}$ and $\partial_{r}$ is reversed, so these regions are
non-static. -- In the second case (figure~\ref{fig:BHHS}c) we have $r_{BH}=
r_{\alpha}$. Here we have a degenerate horizon separating two non-static regions. --
In the last case (figure~\ref{fig:BHHS}d) we have $r_{\alpha}<r_{BH}$, i.e.,
the acceleration horizon becomes the inner horizon. This is the case where
the $\vartheta$ coordinate is restricted by the condition $P( \vartheta ) > 0$.
We have already mentioned that this case is usually considered unphysical.
Therefore, from now on we assume that
\begin{eqnarray}\label{eq:malpha}
0 < m \, , \quad 0 < \alpha < \frac{1}{2m}.
\end{eqnarray}

In this paper we are interested in lensing, assuming that both the light
sources and the observers are in the domain of outer communication,
i.e. in region 2 of figure~1b. Note that a light ray (or any other signal) that
has left this domain can never re-enter. Therefore we need to consider
only null geodesics in the domain of outer communication. Then the
only singularities we may encounter are the conical singularities on
the axis. The string and the strut may be non-transparent or transparent.
In the first case they will block lightlike geodesics and, thus, cast a
shadow. In the second case lightlike geodesics will split into two when
crossing the conic singularity: To see this, we consider a family
of lightlike geodesics whose initial conditions depends continuously
on a real parameter $u$. If the geodesics pass by the axis in positive
$\varphi$ direction for $u<0$ and in negative $\varphi$ direction for
$u>0$, the geodesic with $u=0$ has two different continuations
after crossing the axis.
Note that this is different from the NUT metric which also features a conic
singularity known as the Misner string: In the latter case geodesics are
continuously differentiable when going through the string, without any
splitting; therefore, it is quite natural in the NUT metric to assume that
the string is transparent, as was also argued by Cl{\' e}ment, Gal'tsov and Guenouche
\cite{Clement2015}.

Since both the C-metric and the NUT metric belong to the
family of Pleba{\'n}ski-Demia{\'n}ski spacetimes \cite{Plebanski1976}, there
is also a vacuum solution that combines both parameters.
Griffiths and Podolsk{\' y} \cite{Griffiths2005} have investigated this metric
and shown that there is an ambiguity in the definition of
the NUT parameter. Moreover, in 2006 Chng et al. \cite{Chng2006} found
another vacuum solution, not in the Pleba{\'n}ski-Demia{\'n}ski
class, that describes accelerating black holes with a NUT
parameter and investigated its properties. Podolsk{\' y} and
Vr{\' a}tn{\' y} \cite{Podolsky2020} extended this analysis and showed that
for vanishing NUT charge the metric reduces to the classical
C-metric.

Note that the domain of outer communication is time-symmetric.
On the outer side it is bounded by a future acceleration horizon and
by a past acceleration horizon, whereas on the inner side it is bounded
by a future inner (black hole) horizon and by a past inner (white hole)
horizon. We have already said that for our discussion of lensing
features we have to consider only (sections of) lightlike geodesics
that are confined to the domain of outer communication. However,
if we extend such geodesics into the future they may cross
one of the future horizons, and if we extend them into
the past they may cross one of the past horizons. The \emph{maximal}
analytical extension of the C-metric actually describes two black holes
that are accelerating in opposite directions, with their domains
of outer communication being causally disconnected, see again
Griffiths and Podolsk{\' y} \cite{Griffiths2009}. However, this maximal
extension is of no interest for the purpose of this paper.

\section{Lightlike Geodesics}\label{sec:geo}
In this section we are going to solve the equations of motion for
light rays in the C-metric. In general, the solutions are given in terms of
elliptic integrals and elliptic functions which, for some special cases, simplify
to elementary integrals and elementary functions. While in general there are
different representations and different types of elliptic integrals and
elliptic functions, in this paper we use the elliptic integrals of first and
third kind in their canonical form as well as the Jacobian elliptic functions to
solve the equations of motion. Since these may not be known to all readers we
will define the elliptic integrals used in this paper and show how to use the
Jacobian elliptic functions to solve differential equations in Appendices A and B.

We reduce the equations of motion for lightlike geodesics to first-order form
with the help of constants of motion and we solve them with initial conditions
$(x^{\mu}(\lambda=
\lambda_{i}))=(x_{i}^{\mu})=(t_{i},r_{i},\vartheta_{i},\varphi_{i})$,
where $\lambda_{i}$ is the initial value of the Mino parameter. For the
time being, we leave the initial values arbitrary. For explicit calculations
we will later choose $\lambda _i =0$, which is always possible,
and also $t_i =0$ and $\varphi _i =0$, which is no restriction of
generality because the metric is invariant under time translations and
rotations about the axis of symmetry. We also have to bring to the
reader's attention that although for $\alpha\rightarrow 0$ the geodesics
of the C-metric reduce, of course, to the geodesics of the Schwarzschild
metric, this is not always obvious from the expressions with $\alpha \neq 0$.

\subsection{Equations of motion, photon sphere and photon cone}\label{eq:phc}
The equations of motion for lightlike geodesics are completely integrable.
There are four constants of motion which allow us to rewrite the null geodesic
equation in first order form: the Lagrangian $\mathcal{L}=0$, the energy $E$,
the $z$ component of the angular momentum $L_z$ and the Carter constant $K$.
The resulting first-order equations read
\begin{eqnarray}\label{eq:geot}
\frac{\mathrm{d}t}{\mathrm{d}\lambda} = \frac{r^2E}{Q(r)} \, ,
\end{eqnarray}
\begin{eqnarray}\label{eq:geophi}
\frac{\mathrm{d} \varphi}{\mathrm{d} \lambda} =
\frac{L_z}{P(\vartheta ) \, \mathrm{sin} ^2 \vartheta} \, ,
\end{eqnarray}
\begin{eqnarray}\label{eq:geor}
\left( \frac{\mathrm{d}r}{\mathrm{d} \lambda} \right) ^2 = r^4 E^2 - r^2 Q(r) \, K \, ,
\end{eqnarray}
\begin{eqnarray}\label{eq:geotheta}
\left( \frac{\mathrm{d} \vartheta}{\mathrm{d} \lambda} \right) ^2 =
P ( \vartheta ) \, K  - \frac{L_z^2}{\mathrm{sin} ^2 \vartheta} \, .
\end{eqnarray}
Here $\lambda$ is the Mino parameter \cite{Mino2003}. It is related to an affine parameter $s$ by
\begin{eqnarray}\label{eq:lambda}
\frac{\mathrm{d} \lambda}{\mathrm{d} s}  = \frac{\Omega (r , \vartheta ) ^2}{r^2} \, ,
\end{eqnarray}
cf. e.g. eqs. (16) in Grenzebach et al. \cite{Grenzebach2015a}. Note that $E \neq 0$
for all lightlike geodesics. Without loss of generality, we restrict to geodesics with
$E>0$ throughout this paper. According to (\ref{eq:geot}) this
is tantamount to requiring that the Mino parameter is future-directed with
respect to the $t$ coordinate.

\begin{figure}[ht]
	\includegraphics[width=0.8\textwidth]{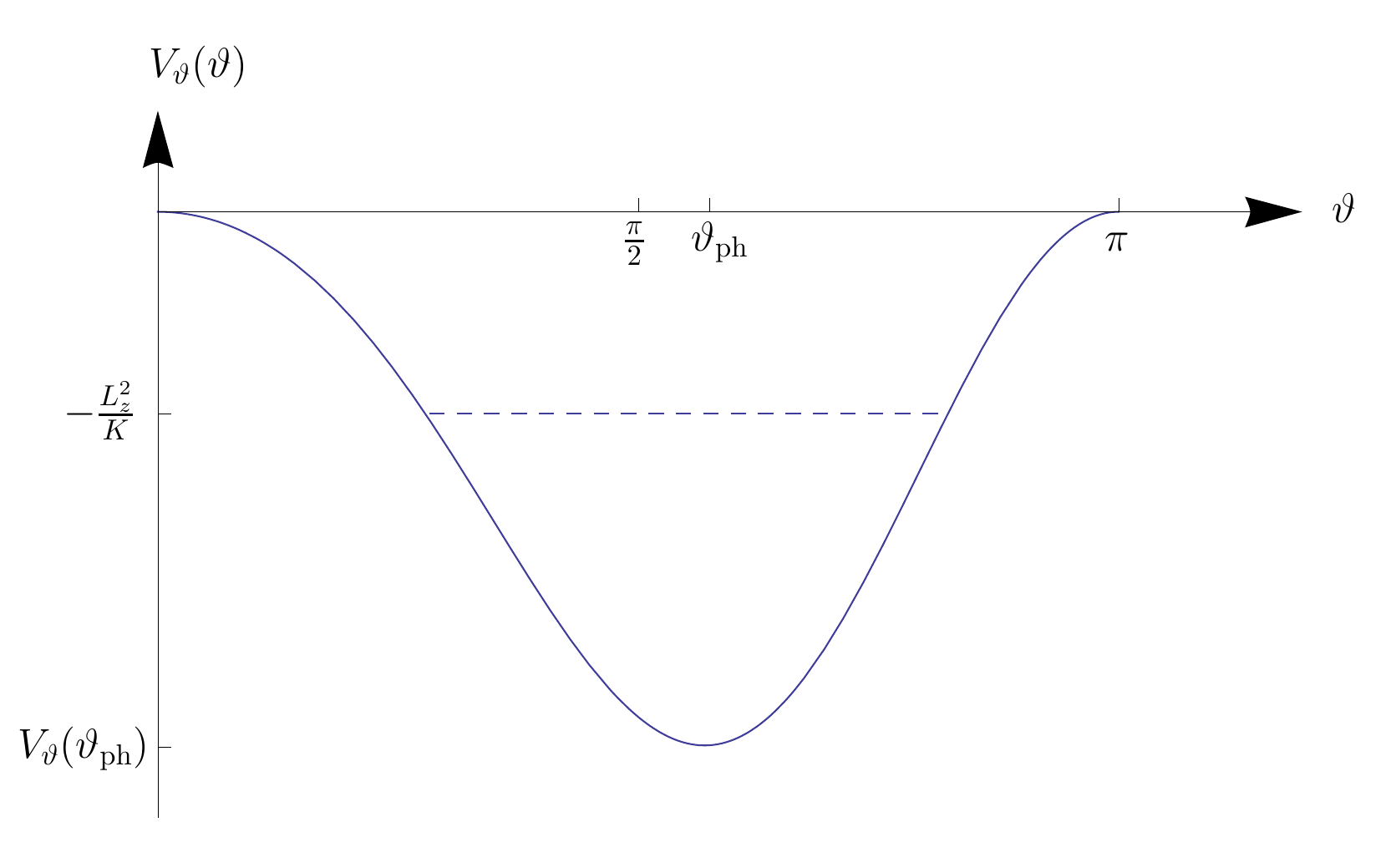}
\caption{The potential $V_{\vartheta} ( \vartheta )$ for latitudinal motion.}
\label{fig:Vtheta}
\end{figure}

From (\ref{eq:geotheta}) we read that $K \ge 0$. Moreover, $K=0$ implies
$d \vartheta / d \lambda =0$ and $L_z=0$; by (\ref{eq:geophi}), the latter
equation implies $d \varphi / d \lambda =0$. This demonstrates that the null
geodesics with $K=0$ are ingoing or outgoing radial light rays. They are the
principal null rays of the C-metric. (The dependence of their $r$ and
$t$ coordinates on the Mino parameter will be given later).
With the case  $K=0$ well understood, we may restrict in the following discussion
to the case $K > 0$.
We first discuss the $\vartheta$ motion. (\ref{eq:geotheta}) may be rewritten as
\begin{eqnarray}\label{eq:thetaE}
\frac{\mathrm{sin} ^2 \vartheta}{K} \, \Big( \frac{\mathrm{d} \vartheta}{\mathrm{d} \lambda} \Big) ^2
+ V_{\vartheta} ( \vartheta ) = \, - \, \frac{L_z^2}{K},
\end{eqnarray}
where
\begin{eqnarray}\label{eq:Vtheta}
V_{\vartheta} ( \vartheta ) = \, - \, P( \vartheta ) \, \mathrm{sin} ^2 \vartheta
= \, - \, (1-2 \, \alpha \, m \, \mathrm{cos} \, \vartheta ) \, \mathrm{sin} ^2 \vartheta
\, .
\end{eqnarray}
(\ref{eq:thetaE}) demonstrates that null geodesics with constants of motion
$L_z$ and $K$ may exist only in the region where $V_{\vartheta} ( \vartheta )
 < -L_z^2/K$, see figure~\ref{fig:Vtheta}. The potential $V_{\vartheta}(\vartheta)$ has a
 minimum at $\vartheta = \vartheta _{\mathrm{ph}}$ given by
\begin{eqnarray}\label{eq:thetaph}
\mathrm{cos} \, \vartheta _{\mathrm{ph}} =
\frac{- \, 2 \, \alpha \, m}{1+
\sqrt{1+12 \, \alpha ^2 m^2}}
\, .
\end{eqnarray}
The allowed values for $L_z$ and $K$ are
\begin{eqnarray}\label{eq:LzKbounds}
0 \, \ge \, - \, \frac{L_z^2}{K} \, \ge \, V_{\vartheta} ( \vartheta _{\mathrm{ph}} )
\, .
\end{eqnarray}
Null geodesics with $-L_z^2 /K = V_{\vartheta} ( \vartheta _{\mathrm{ph}} )$
are completely contained in the cone $\vartheta = \vartheta _{\mathrm{ph}}$.
Hereafter we will refer to it as the \emph{photon cone}.
Null geodesics with $L_z =0$ meet the axis. All other null
geodesics oscillate between a minimum value, $0 < \vartheta _{\mathrm{min}} <
\vartheta _{\mathrm{ph}}$, and a maximum value, $\vartheta _{\mathrm{ph}} <
\vartheta _{\mathrm{max}} < \pi$.
Note that $\vartheta _{\mathrm{ph}} \to \pi /2$
if $\alpha \to 0$ and $\vartheta _{\mathrm{ph}} \to \mathrm{arccos} (-1/3) \approx 1.911$ if $\alpha \to 1 / (2 \, m)$.

\begin{figure}[ht]
	\includegraphics[width=0.7\textwidth]{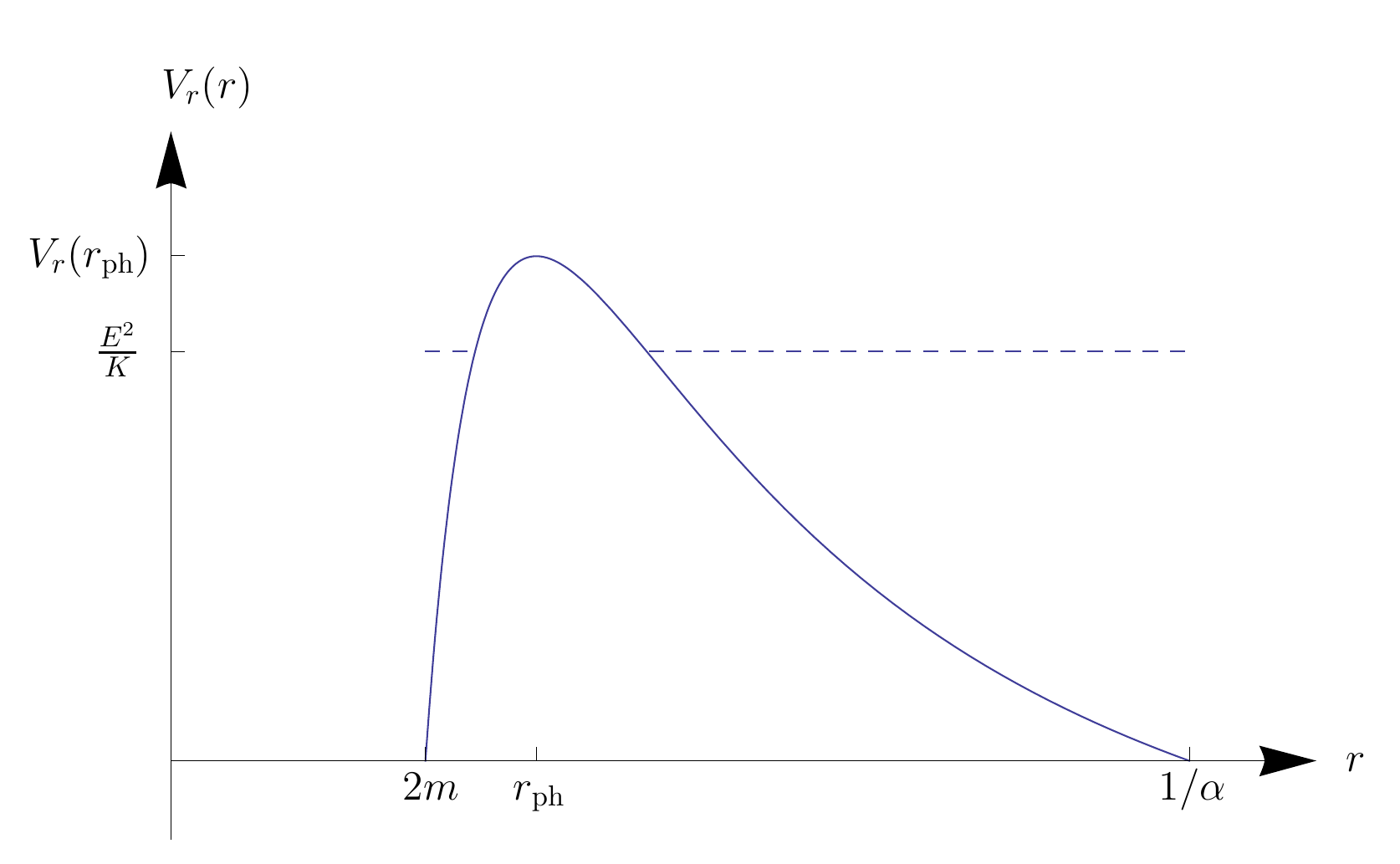}
\caption{The potential $V_r (r )$ for radial motion.}
\label{fig:Vr}
\end{figure}

We now turn to the $r$ motion. Eq.~(\ref{eq:geor}) may be rewritten as
\begin{eqnarray}\label{eq:rE}
\frac{1}{r^4K} \, \Big( \frac{\mathrm{d} r}{\mathrm{d} \lambda} \Big) ^2  + V_r(r)
= \frac{E^2}{K},
\end{eqnarray}
where
\begin{eqnarray}\label{eq:Vr}
V_r(r) = \frac{Q(r)}{r^2} = \Big( \frac{1}{r^2} - \alpha ^2 \Big)
\Big( 1 - \frac{2m}{r} \Big) \, .
\end{eqnarray}
Null geodesics with constants of motion $E$ and $K$ exist in the domain
where $V_r(r) \le E^2/K$, see figure~\ref{fig:Vr}. The potential $V_r(r)$ has a
maximum at $r = r_{\mathrm{ph}}$ where
\begin{eqnarray}\label{eq:rph}
r _{\mathrm{ph}} =
\frac{6 \, m}{1 + \sqrt{1 + 12 \, \alpha ^2 m^2 }}
\, .
\end{eqnarray}

Note that $r _{\mathrm{ph}} \to 3 \, m$ if $\alpha \to 0$
and $r _{\mathrm{ph}} \to 2 \, m$ if $\alpha \to 1 / (2 \, m)$. The sphere
$r = r_{\mathrm{ph}}$ is called the \emph{photon sphere} or the \emph{light sphere}.
If a null geodesic is tangential to this sphere at one point, then it is
completely contained in this sphere.

Null geodesics with
$E^2/K < V_r ( r_{\mathrm{ph}} )$ have exactly one extremum of the $r$ motion
which is either a maximum at a value $2 \, m < r _{\mathrm{max}} < r _{\mathrm{ph}}$
or a minimum at a value $r _{\mathrm{ph}} < r _{\mathrm{min}} < 1 / \alpha$.
Null geodesics with $E^2/K > V_r ( r_{\mathrm{ph}} )$ have no extremum of
the $r$ motion. Null geodesics with $E^2/K = V_r ( r_{\mathrm{ph}} )$ are
either completely contained in the photon sphere or they spiral asymptotically
towards the photon sphere. Of all the null geodesics that fill the photon
sphere, only the ones through the axis lie on halves of great circles. Whenever
a null geodesic crosses the axis the $\varphi$ coordinate of the half of the
great circle changes and generally the halves before and after crossing the axis
do not form a closed great circle (for more details see Section 3.4). At the
intersection of $r= r_{\mathrm{ph}}$ and $\vartheta = \vartheta _{\mathrm{ph}}$
there is a circular null geodesic, but it is not a great circle. All the
other null geodesics on the photon sphere are non-circular.

The photon sphere is of crucial relevance for lensing, in particular for
the formation of the shadow. This was discussed in detail, for a class
of spacetimes that includes the C-metric, by Grenzebach
et al. \cite{Grenzebach2015a,Grenzebach2016}. It should be mentioned
that the photon sphere is not a ``round sphere'', i.e., its intrinsic
geometry is not that of the standard sphere in Euclidean 3-space, see
Gibbons and Warnick \cite{Gibbons2016}. The photon cone appeared
in the literature only recently: Alrais Alawadi et al. \cite{AlraisAlawadi2021}
found that there is a circular lightlike geodesic at the intersection of
this cone with the photon sphere. (Note that their $\alpha$ is our
$- \alpha$.) We believe that our characterisation of the photon cone in
terms of the potential $V_{\vartheta}(\vartheta)$ is
particularly transparent and gives additional useful information.

Having discussed some qualitative features of lightlike
geodesics with the help of the potentials (\ref{eq:Vtheta}) and
(\ref{eq:Vr}), we will now explicitly solve the equations of motion
(\ref{eq:geot})-(\ref{eq:geotheta}). Generically, this will be done
in terms of elliptic integrals and elliptic functions. For some
special geodesics the solutions reduce to elementary functions.
For each of the four equations we will first characterise the special
cases and then the generic ones.

\subsection{The $r$ motion}
\subsubsection{Types of motion}
The fourth-order polynomial on the right-hand side of  (\ref{eq:geor}) has always
four roots $r_1$, $r_2$, $r_3$ and $r_4$ in the complex plane which
can be determined, e.g., with Cardano's
method. Real roots in the interval between $r_{BH}$ and $r_{\alpha}$ give
turning points of the geodesic. Note that there is always a root at $r=0$ and another
one at some real value $r \le 0$. The remaining two roots are either real and
non-negative or non-real and complex conjugate to each other. We have to
distinguish the special cases where two or more roots coincide from the generic
cases where the four roots are distinct. This will give us four qualitatively
different types of $r$ motion. The roots depend, of course, on the constants
of motion. Therefore, the four different types of $r$ motion can be characterised
in terms of the constants of motion. For this purpose it is convenient to define
a new constant of motion $K_{E}=\frac{K}{E^2}$ (recall that $E \neq 0$ for all
lightlike geodesics and that we choose $E>0$) and to introduce the abbreviation
\begin{eqnarray}\label{eq:Kph}
K_{ph}=\frac{r_{ph}^2}{Q(r_{ph})},
\end{eqnarray}
where $r_{ph}$ is the radius of the photon sphere.
We can then characterise the four types of $r$ motion as follows:
\begin{enumerate}
\item $K_{E}=0$: These are the principal null geodesics. The polynomial has
four real roots all of which coincide, $r_{1}=r_{2}=r_{3}=r_{4}=0$.
\item $K_{E}=K_{ph}$: These are null geodesics on the photon sphere or asymptotically approaching
the photon sphere in the future or in the past. The polynomial has four real roots two of which
coincide; we label them such that $r_{4}<r_{3}=0<r_{2}=r_{1}=r_{ph}$.
\item $K_{ph}<K_{E}$: In the domain $r_{BH}<r<r_{ph}$ these are geodesics with a maximum
at $r_{max}=r_{2}$ and in the domain $r_{ph}<r<r_{\alpha}$ these are geodesics with a minimum
at $r_{min}=r_{1}$. The polynomial has four distinct real roots which we label such that
$r_{4}<r_{3}=0<r_{2}<r_{1}$.
\item $0<K_{E}<K_{ph}$: These are null geodesics without turning points. The polynomial
has four distinct roots two of which are real and two are complex conjugate; we label them such
that $0=r_{1}>r_{2}$, $r_{3}=R_{3}+iR_{4}$ and $r_{4}=R_{3}-iR_{4}$ with $R_3$
real and $R_4$ positive.
\end{enumerate}
We now discuss each of the four types of $r$ motion separately.

\subsubsection{Principal null geodesics:}
For principal null geodesics we have $K=0$. In this case (\ref{eq:geor}) reduces to
\begin{eqnarray}\label{eq:georrad}
\left( \frac{\mathrm{d}r}{\mathrm{d} \lambda} \right) ^2 = r^4 E^2.
\end{eqnarray}
With the initial condition $r(\lambda_{i})=r_{i}$ and indicating the sign of
$dr/d \lambda$ at $\lambda _{i}$ by $i_{r_{i}}$ we get as solution
\begin{eqnarray}\label{eq:georadsol}
r(\lambda)=\frac{r_{i}}{1-i_{r_{i}}r_{i}E(\lambda-\lambda_{i})}.
\end{eqnarray}

\subsubsection{Geodesics with $K_{E}=K_{ph}$:}
When we have $K_{E}=K_{ph}$ (\ref{eq:geor}) has a double zero at $r=r_{ph}$.
Therefore we have either null geodesics that are trapped on the photon sphere or
null geodesics that asymptotically come from or go to the photon sphere. In the
former case the solution is trivial, $r(\lambda)=r_{i}=r_{ph}$. In the latter case
we substitute
\begin{eqnarray}\label{eq:rsub1}
r=\frac{6mK}{12y+K}
\end{eqnarray}
in (\ref{eq:geor}) and obtain
\begin{eqnarray}\label{eq:geoy}
\left( \frac{\mathrm{d}y}{\mathrm{d} \lambda} \right) ^2=4y^3-g_{2,r}y-g_{3,r} \, .
\end{eqnarray}
Here, $g_{2,r}$ and $g_{3,r}$ are constants of motion that can be expressed in
terms of $m$, $\alpha$, $E$ and $K$. The explicit form of these expressions need
not be given here. Expressing the polynomial in terms of its roots we obtain:
\begin{eqnarray}\label{eq:geoyroots}
\left( \frac{\mathrm{d}y}{\mathrm{d} \lambda} \right) ^2=4(y-y_{ph})^2(y-y_{1})
\, .
\end{eqnarray}
$y_{ph}$ and $y_{1}$ are related to $r_{ph}$ and $r_{4}$ via (\ref{eq:rsub1}). Using
this transformation it is easy to show that we have $y_{ph}<y$ for $r_{BH}<r<r_{ph}$,
and $y<y_{ph}$ for $r_{ph}<r<r_{\alpha}$. Similarly one can easily show that
$y_{1}<y_{ph}$ and $y_{1}<y$.

Now separation of variables and integrating from $y(\lambda_{i})=y_{i}$ to $y(\lambda)$
leads to
\begin{eqnarray}\label{eq:geoyint}
\lambda-\lambda_{i}=-\frac{i_{r_{i}}}{2}\int_{y_{i}}^{y}\frac{\mathrm{d}y'}{\sqrt{(y'-y_{ph})^2(y'-y_{1})}}.
\end{eqnarray}
Here we have to distinguish null geodesics between the black hole horizon and the photon sphere
from null geodesics between the photon sphere and the acceleration horizon. In the former case
we have $y_{ph}<y$ ($r_{BH}<r<r_{ph}$) and rewrite the right-hand side using $I_{1}$
in (\ref{eq:int}). In the latter case we have $y<y_{ph}$ ($r_{ph}<r<r_{\alpha}$)
and rewrite the right-hand side using $I_{2}$ in (\ref{eq:int}).

For $r_{BH}<r<r_{ph}$ we now integrate following the steps in Appendix A.1 and obtain
(\ref{eq:intsol1}). After inserting in (\ref{eq:geoyint}) and solving for
$r$ we obtain
\begin{eqnarray}\label{eq:georrphminus}
\hspace*{-2.25cm}r(\lambda)=\frac{r_{ph}r_{4}}{r_{ph}-(r_{ph}-r_{4})\coth^2\left(\mathrm{arcoth}\left(\sqrt{\frac{r_{ph}(r_{i}-r_{4})}{r_{i}(r_{ph}-r_{4})}}\right)+i_{r_{i}}\sqrt{\frac{mK(r_{4}-r_{ph})}{2r_{ph}r_{4}}}\left(\lambda-\lambda_{i}\right)\right)}.
\end{eqnarray}
Following the analogous steps for $r_{ph}<r<r_{\alpha}$ we get (\ref{eq:intsol2}).
This time solving for $r$ leads to
\begin{eqnarray}\label{eq:georrphplus}
\hspace*{-2.25cm}r(\lambda)=\frac{r_{ph}r_{4}}{r_{ph}-(r_{ph}-r_{4})\tanh^2\left(\mathrm{artanh}\left(\sqrt{\frac{r_{ph}(r_{i}-r_{4})}{r_{i}(r_{ph}-r_{4})}}\right)-i_{r_{i}}\sqrt{\frac{mK(r_{4}-r_{ph})}{2r_{ph}r_{4}}}\left(\lambda-\lambda_{i}\right)\right)}.
\end{eqnarray}

\subsubsection{Geodesics with $K_{ph}<K_{E}$:}
Again we have to distinguish null geodesics between the black hole horizon $r_{BH}$ and the
photon sphere $r_{ph}$ from null geodesics between the photon sphere $r_{ph}$ and the
accreleration horizon $r_{\alpha}$. In the former case the
geodesic has a turning point at $r_{max}=r_{2}$. In this case we use the
transformation \cite{Hancock1917,Gralla2020}
\begin{eqnarray}\label{eq:georsub2}
r=r_{1}-\frac{r_{1}(r_{1}-r_{2})}{r_{1}-r_{2}\sin^2\chi_{r}}
\end{eqnarray}
to put (\ref{eq:geor}) into the Legendre form (\ref{eq:Legendrediff}). Now we
follow the steps described in Appendix B to obtain the solution for $r(\lambda)$
in terms of Jacobi's $\mathrm{sn}$ function:
\begin{eqnarray}\label{eq:georsolmax}
\hspace*{-1.0cm}r(\lambda)=r_{1}-\frac{r_{1}(r_{1}-r_{2})}{r_{1}-r_{2}\mathrm{sn}^2\left(i_{r_{i}}\frac{\sqrt{(E^2+\alpha^2 K)r_{1}(r_{2}-r_{4})}}{2}(\lambda_{i}-\lambda)+\lambda_{r_{i},k_{1}},k_{1}\right)}.
\end{eqnarray}
Here, $k_{1}$ and $\lambda_{r_{i},k_{1}}$ are given by the following two relations:
\begin{eqnarray}\label{eq:modulus1}
k_{1}=\frac{r_{2}(r_{1}-r_{4})}{r_{1}(r_{2}-r_{4})},
\end{eqnarray}
\begin{eqnarray}\label{eq:initial1}
\lambda_{r_{i},k_{1}}=F_{L}(\chi_{r,i},k_{1}), \chi_{r,i}=\arcsin\left(\sqrt{\frac{(r_{2}-r_{i})r_{1}}{(r_{1}-r_{i})r_{2}}}\right).
\end{eqnarray}
Analogously for null geodesics between the photon sphere $r_{ph}$ and the acceleration horizon
$r_{\alpha}$ we use the transformation \cite{Hancock1917,Gralla2020}
\begin{eqnarray}\label{eq:georsub3}
r=r_{2}+\frac{(r_{1}-r_{2})(r_{2}-r_{4})}{r_{2}-r_{4}-(r_{1}-r_{4})\sin^2\chi_{r}}
\end{eqnarray}
to put (\ref{eq:geor}) into the Legendre form (\ref{eq:Legendrediff}). Again
we follow the procedure described in Appendix B and obtain the solution for
$r(\lambda)$, which again involves Jacobi's $\mathrm{sn}$ function and this time
reads
\begin{eqnarray}\label{eq:georsolmin}
\hspace*{-2.0cm}r(\lambda)=r_{2}+\frac{(r_{1}-r_{2})(r_{2}-r_{4})}{r_{2}-r_{4}-(r_{1}-r_{4})\mathrm{sn}^2\left(i_{r_{i}}\frac{\sqrt{(E^2+\alpha^2 K)r_{1}(r_{2}-r_{4})}}{2}(\lambda-\lambda_{i})+\lambda_{r_{i},k_{1}},k_{1}\right)},
\end{eqnarray}
where
\begin{eqnarray}\label{eq:initial2}
\lambda_{r_{i},k_{1}}=F_{L}(\chi_{r,i},k_{1}), \chi_{r,i}=\arcsin\left(\sqrt{\frac{(r_{i}-r_{1})(r_{2}-r_{4})}{(r_{i}-r_{2})(r_{1}-r_{4})}}\right).
\end{eqnarray}

\subsubsection{Geodesics with $0<K_{E}<K_{ph}$:}
Null geodesics with $0<K_{E}<K_{ph}$ do not possess turning points. Therefore in
this case we only need one transformation to put (\ref{eq:geor}) into the Legendre
form. Using the real root $r_{2}$ and the real and imaginary parts $R_{3}$ and $R_{4}$
of the complex conjugate roots $r_{3}$ and $r_{4}$ we first define two new constants
of motion $R$ and $\bar{R}$ given by
\begin{eqnarray}\label{eq:LegendreConstants}
R=\sqrt{R_{3}^2+R_{4}^2},~\mathrm{and}~\bar{R}=\sqrt{(R_{3}-r_{2})^2+R_{4}^2}.
\end{eqnarray}
Now we use the transformation \cite{Hancock1917,Gralla2020}
\begin{eqnarray}\label{eq:georsub4}
r=\frac{r_{2}R(\cos\chi_{r}-1)}{\bar{R}-R+(\bar{R}+R)\cos\chi_{r}},
\end{eqnarray}
to put (\ref{eq:geor}) in the Legendre form (\ref{eq:Legendrediff}). Again we
follow the steps described in Appendix B. This time the solution for $r(\lambda)$
is given in terms of Jacobi's $\mathrm{cn}$ function. It reads
\begin{eqnarray}\label{eq:georsolnoturn}
\hspace*{-1.5cm}r(\lambda)=\frac{r_{2}R\left(\mathrm{cn}\left(i_{r_{i}}\sqrt{(E^2+\alpha^2 K)R\bar{R}}(\lambda-\lambda_{i})+\lambda_{r_{i},k_{2}},k_{2}\right)-1\right)}{\bar{R}-R+(\bar{R}+R)\mathrm{cn}\left(i_{r_{i}}\sqrt{(E^2+\alpha^2 K)R\bar{R}}(\lambda-\lambda_{i})+\lambda_{r_{i},k_{2}},k_{2}\right)},
\end{eqnarray}
where $k_{2}$ and $\lambda_{r_{i},k_{2}}$ are given by
\begin{eqnarray}\label{eq:modulus2}
k_{2}=\frac{(R+\bar{R})^2-r_{2}^2}{4R\bar{R}},
\end{eqnarray}
\begin{eqnarray}\label{eq:initial3}
\lambda_{r_{i},k_{2}}=F_{L}(\chi_{r,i},k_{2})~\mathrm{and}~\chi_{r,i}=\arccos\left(\frac{(r_{i}-r_{2})R-r_{i}\bar{R}}{(r_{i}-r_{2})R+r_{i}\bar{R}}\right).
\end{eqnarray}

\subsection{The $\vartheta$ Motion}
For discussing the $\vartheta$ motion it is convenient to rewrite  (\ref{eq:geotheta})
as a differential equation for  $x=\cos\vartheta$,
\begin{eqnarray}\label{eq:geocostheta}
\left( \frac{\mathrm{d} x}{\mathrm{d} \lambda} \right) ^2 = (1-x^2)(1-2\alpha mx)K-L_{z}^2.
\end{eqnarray}
We immediately see that when we have $K=L_{z}^2=0$ the right-hand side vanishes identically.
In all other cases the third-order polynomial on the right-hand side must have three real roots
because, as discussed in Section \ref{eq:phc}, we always have two turning points. Using this
information we can now distinguish the following three types of $\vartheta$ motion:
\begin{enumerate}
\item $K=L_{z}^2=0$: These are again the principal null geodesics. The right-hand side of
(\ref{eq:geocostheta}) vanishes identically.
\item $K=\frac{L_{z}^2}{\sin^2\vartheta_{ph}P(\vartheta_{ph})} \neq 0$: These are null
geodesics on the photon cone. Two of the three real roots coincide; we label them such that
$x_{1}>x_{2}=x_{3}=\cos\vartheta_{ph}$.
\item  $K \neq \frac{L_{z}^2}{\sin^2\vartheta_{ph}P(\vartheta_{ph})}$: These are
null geodesics where the $\vartheta$ coordinate oscillates between a minimum and a maximum.
The three real roots are distinct; we label them such that $x_{3}<x_{2}<x_{1}$ where
$x_{2}=\cos\vartheta_{\mathrm{min}}$ and $x_{3}=\cos\vartheta_{\mathrm{max}}$.
\end{enumerate}
For (i) and (ii) (\ref{eq:geocostheta}) is easy to solve. In both cases we have
$\frac{\mathrm{d}x}{\mathrm{d}\lambda}=\frac{\mathrm{d}^2x}{\mathrm{d}\lambda^2}=0$
and thus $x=\cos\vartheta=\mathrm{const}$. Thus in both cases the solution is
$\vartheta(\lambda)=\vartheta_{i}$.

Now we turn to case (iii). Here we first transform (\ref{eq:geocostheta}) to the
Legendre form (\ref{eq:Legendrediff}) using the following coordinate transformation \cite{Hancock1917}:
\begin{eqnarray}\label{eq:geothetasub1}
x=\cos\vartheta=(x_{2}-x_{3})\sin^2\chi_{\vartheta}+x_{3} \, .
\end{eqnarray}
As for the $r$ motion we follow the steps in Appendix B to solve the differential
equation and obtain the solution for $\vartheta(\lambda)$ in terms of Jacobi's
$\mathrm{sn}$ function:
\begin{eqnarray}\label{eq:geothetasolplus}
\hspace*{-2.0cm}\vartheta(\lambda)=\arccos\left((x_{2}-x_{3})\mathrm{sn}^2\left(i_{\vartheta_{i}}\sqrt{\frac{\alpha mK(x_{1}-x_{3})}{2}}(\lambda_{i}-\lambda)+\lambda_{\vartheta_{i},k_{3}},k_{3}\right)+x_{3}\right) \, .
\end{eqnarray}
Here, $i_{\vartheta_{i}}$ denotes the sign of $\mathrm{d} \vartheta / \mathrm{d} \lambda $
at $\lambda_{i}$. The square of the elliptic modulus $k_{3}$ and
$\lambda_{\vartheta_{i},k_{3}}$ are given by
\begin{eqnarray}\label{eq:modulus3}
\hspace*{-2.0cm}k_{3}=\frac{x_{2}-x_{3}}{x_{1}-x_{3}}, \mathrm{and}~\lambda_{\vartheta_{i},k_{3}}=F_{L}(\chi_{\vartheta,i},k_{3}), \chi_{\vartheta,i}=\arcsin\left(\sqrt{\frac{\cos\vartheta_{i}-x_{3}}{x_{2}-x_{3}}}\right).
\end{eqnarray}

\subsection{The $\varphi$ motion}
As the right-hand side of (\ref{eq:geophi}) is known once we have calculated
$\vartheta  ( \lambda )$, the $\varphi$ motion is determimed by the $\vartheta$
motion. Therefore the different types of $\varphi$ motion correspond to the
different types of the $\vartheta$ motion:
\begin{enumerate}
\item $K=L_{z}^2=0$: These are  the principal null geodesics for which the
right-hand side of (\ref{eq:geophi}) vanishes. So for these geodesics we have
$\varphi(\lambda)=\varphi_{i}$.
\item $K=\frac{L_{z}^2}{\sin^2\vartheta_{ph}P(\vartheta_{ph})} \neq 0$: These are null
geodesics on the photon cone. For them the right-hand side of (\ref{eq:geophi}) is
constant, so the solution is
\begin{eqnarray}\label{eq:geophisolphoton}
\varphi(\lambda)=\varphi_{i}+\frac{L_{z}(\lambda-\lambda_{i})}{P(\vartheta_{ph})\sin^2\vartheta_{ph}} \, .
\end{eqnarray}
\item  $K \neq \frac{L_{z}^2}{\sin^2\vartheta_{ph}P(\vartheta_{ph})}$: Here the
geodesics with $L_z=0$ have to be considered separately. These are geodesics which
meet the axis $\mathrm{sin} \, \vartheta =0$. If the string and the strut
are non-transparent, then these geodesics are blocked at the axis. Otherwise they
split into two geodesics which can be calculated by continuously extending sequences
of geodesics with positive and negative $L_z$, respectively.
For the null geodesics with $L_z \neq 0$ we proceed similarly to the
$\vartheta$ motion. We first rewrite the right-hand side of (\ref{eq:geophi}) in terms of
the variable $x=\cos\vartheta$:
\begin{eqnarray}\label{eq:geophicos}
\frac{\mathrm{d}\varphi}{\mathrm{d}\lambda} = \frac{L_z}{(1-x^2)(1-2\alpha m x)} \, .
\end{eqnarray}
After dividing (\ref{eq:geophicos}) by $\pm dx/d \lambda$ and substituting for the latter
from (\ref{eq:geocostheta}) the resulting equation can be integrated:
\begin{eqnarray}\label{eq:geophicosint}
\hspace*{-1.75cm}\varphi(\lambda)=\varphi_{i}+\int_{\cos\vartheta_{i}...}^{...\cos\vartheta(\lambda)}\frac{L_{z}\mathrm{d}x}{(1-x^2)(1-2\alpha mx) \sqrt{(1-x^2)(1-2 \alpha m x) K-L_{z}^2}}.
\end{eqnarray}
Here the dots indicate that we have to split the integral at each turning point, choosing the
sign of $\pm dx/d \lambda$ such that $L_z^{-1} \mathrm{d} \varphi /\mathrm{d} \lambda$
is always positive.
Next we perform a partial fraction decomposition of $(1-x^2)^{-1}(1-2\alpha m x)^{-1}$. Then we
use (\ref{eq:geothetasub1}) and express (\ref{eq:geophicosint}) by elliptic
integrals of the third kind (\ref{eq:EllipticIntegrals}).

\end{enumerate}

\subsection{The time coordinate $t$}\label{subsec:time}
For calculating the time coordinate $t$ as  a function of $\lambda$ we have to distinguish
the same four types of motion as for the radius coordinate $r$. In all cases but one we start
with (\ref{eq:geot}) and divide by $\pm dr/d \lambda$ as given by (\ref{eq:geor}). Next
we integrate and obtain
\begin{eqnarray}\label{eq:geotint}
\hspace*{-1.5cm}t(\lambda)=t_{i}+
\int_{r_{i}...}^{...r(\lambda)}\frac{Er'^2\mathrm{d}r'}{Q(r')\sqrt{r'^4E^2-r'^2Q(r')K}}.
\end{eqnarray}
Here the dots indicate that we have to split the integral at a turning point
and to choose the sign of $\pm dr/d \lambda$ such that $dt / d \lambda$ is
always positive.

\subsubsection{Principal null geodesics}
We first set $K=0$ in (\ref{eq:geotint}). Then we re-arrange the term $Q(r)^{-1}$
such that only terms with $r$ in the denominator remain and perform a partial
fraction decomposition. This leaves us with
three elementary integrals that can be easily evaluated,
\begin{eqnarray}\label{eq:geotsolradial}
\hspace*{-2.25cm}t(\lambda)=&t_{i}+i_{r_{i}}\left(\frac{2m}{1-4\alpha^2m^2}\ln\left(\frac{r(\lambda)-2m}{r_{i}-2m}\right)+\frac{1}{2\alpha (1-2\alpha m)}\ln\left(\frac{1-\alpha r_{i}}{1-\alpha r(\lambda)}\right)\right.\\
\hspace*{-2.25cm}&\left.+\frac{1}{2\alpha (1+2\alpha m)}\ln\left(\frac{1+\alpha r(\lambda)}{1+\alpha r_{i}}\right)\right) \, . \nonumber
\end{eqnarray}
Note that this is the same expression as derived by Lim \cite{Lim2014}. The only difference
is that we expressed the integration constant in terms of the initial conditions.

\subsubsection{Geodesics with $K_{E}=K_{ph}$}
The first case we have to consider are null geodesics on the photon sphere. For these
geodesics we have $r=r_{ph}=\mathrm{const}$. In this case inserting
$r_{ph}$ into (\ref{eq:geot}) and integrating over $\lambda$ yields:
\begin{eqnarray}\label{eq:geotsolphoton1}
t(\lambda)=t_{i}+\frac{r_{ph}^2E(\lambda-\lambda_{i})}{Q(r_{ph})} \, .
\end{eqnarray}
In the second case we have null geodesics asymptotically coming from or going to
the photon sphere. Here, we first substitute (\ref{eq:rsub1}) into (\ref{eq:geotint})
and perform a partial fraction decomposition. Now we split the integral into
four different terms. After expressing (\ref{eq:geotint}) using the integrals from
Appendix A.1 we find:
\begin{eqnarray}\label{eq:geotsolphoton2}
t(\lambda)=&t_{i}\pm i_{r_{i}}\left(\frac{2r_{ph}m^2E I_{ph,1}}{(1-4\alpha^2m^2)(r_{ph}-2m)}-\frac{r_{ph}E I_{ph,2}}{4\alpha(1-2\alpha m)(1-\alpha r_{ph})}\right.\\
&\left.+ \frac{r_{ph}E I_{ph,3}}{4\alpha(1+2\alpha m)(1+\alpha r_{ph})}\mp\frac{r_{ph}^3E I_{ph,4\pm}}{2(r_{ph}-2m)(1-\alpha^2 r_{ph}^2)}\right). \nonumber
\end{eqnarray}
Here, the upper sign is valid for null geodesics between the photon sphere and the acceleration
horizon, while the lower sign is valid for null geodesics between the black hole horizon
and the photon sphere. The integrals $I_{ph,1}$-$I_{ph,4\pm}$ in (\ref{eq:geotsolphoton2})
are given by (\ref{eq:intsol1}) and (\ref{eq:intsol2}) in Appendix A.1.

\subsubsection{Geodesics with $K_{ph}<K_{E}$}
Here we proceed as for principal null geodesics. We first rewrite (\ref{eq:geotint})
such that it only contains $r$ in the denominator. Then we perform a partial fraction
decomposition. For null geodesics between the photon sphere and the acceleration horizon
we now use (\ref{eq:georsub3}) to rewrite (\ref{eq:geotint}) using elliptic integrals
of first and third kind. Analogously for null geodesics between the black hole horizon
and the photon sphere we use (\ref{eq:georsub2}) to rewrite (\ref{eq:geotint}) using
elliptic integrals of first and third kind. In the former case the parameter of the
elliptic integral of third kind is $>1$ for the term diverging at the acceleration
horizon while in the latter case it is $>1$ for the term diverging at the black hole
horizon. However, in both cases we chose the coordinate transformations such that
we never integrate over these divergences.

\subsubsection{Geodesics with $0<K_{E}<K_{ph}$}
The first steps are the same as for null geodesics with $K_{ph}<K_{E}$. After the
partial fraction decomposition we use (\ref{eq:georsub4}) and rewrite (\ref{eq:geotint})
using elliptic integrals of first kind and the non-standard elliptic integral
$J(\chi_{i},\chi,k,n)$. The latter we now rewrite as (\ref{eq:ellipticJint})
using elliptic integrals of third kind and elementary functions as demonstrated
in Appendix A.2. Note that for the term diverging at the black hole horizon we
always integrate over the divergence and thus for this term we have to rewrite
the elliptic integrals of third kind using (\ref{eq:Elliptic3div}).

\subsection{Calculation and visualisation of the analytically determined null geodesics}\label{subsec:examples}
For the visualisation of the null geodesics we implemented all solutions in the
programming language Julia \cite{Bezanson2017}. Because it is a common convention
that general null geodesics are future-directed we will limit our description of the
calculation procedure to future-directed null geodesics. Whenever differences to past-directed
null geodesics occur -- which we have to use when we turn to gravitational lensing in Section
\ref{sec:lensing} -- we write them in brackets directly behind the expression for
future-directed null geodesics. For most of the calculations the procedure is straight
forward and can be applied to future-directed and past-directed null geodesics
alike. We set $\lambda_{i}=0$, $t_{i}=0$ and $\varphi_{i}=0$. For any choice of
$r_i$, $\vartheta _i$, $E$, $L_z$ and $K$ we integrate the equations of motion
from $\lambda _i =0$ up to some value $\lambda _f >0$ for future-directed null geodesics
and $\lambda _f <0$ for past-directed null geodesics. The formulas derived in the preceding
sections give us directly $r(\lambda _f)$ and $\vartheta(\lambda _f)$. For calculating
$t(\lambda _f)$ with the help of (\ref{eq:geotint}) we need to know the turning
points of the $r$ motion between $\lambda _i =0$ and $\lambda _f$, and for calculating
$\varphi(\lambda _f)$ with the help of (\ref{eq:geophicosint}) we need to know the
turning points of the $\vartheta$ motion between $\lambda _i =0$ and $\lambda _f$.
We know that the $r$ motion can have only one turning point; between $\lambda _i =0$
and $\lambda _f >0$ (for past-directed null geodesics: $\lambda _f <0$) such a turning point is necessarily a maximum
for initially outgoing geodesics, given by (\ref{eq:georsolmax}), while it is necessarily a
minimum for initially ingoing geodesics, given by (\ref{eq:georsolmin}). By contrast,
the $\vartheta$ motion can potentially have arbitrarily many turning points.

For determining $t ( \lambda _f )$ we proceed in the following way. For an initially
outgoing geodesic with a maximum of the $r$ motion we have to check whether
this maximum $r_{\mathrm{max}}$ comes before $\lambda _f$. To that end we
solve  (\ref{eq:geor}) for $\mathrm{d} \lambda /dr$ and integrate from $r_i$ to
$r_{\mathrm{max}}$. This gives us the Mino parameter $\lambda _{\mathrm{max}}$
at the maximum as an elliptic integral of the first kind. If $\lambda _{\mathrm{max}} <
\lambda _f$ (for past-directed null geodesics: $\lambda _{\mathrm{max}} >\lambda _f$), we have to split the integral
(\ref{eq:geotint}) into two sections, from $\lambda _i$ to $\lambda _{\mathrm{max}}$
and from $\lambda _{\mathrm{max}}$ to $\lambda _f$; otherwise, we just have to
integrate from $\lambda _i$ to $\lambda _f$. For an initially ingoing geodesic
with a minimum of the $r$ motion the procedure is analogous.

\begin{figure}[ht]
	\includegraphics[width=0.8\textwidth]{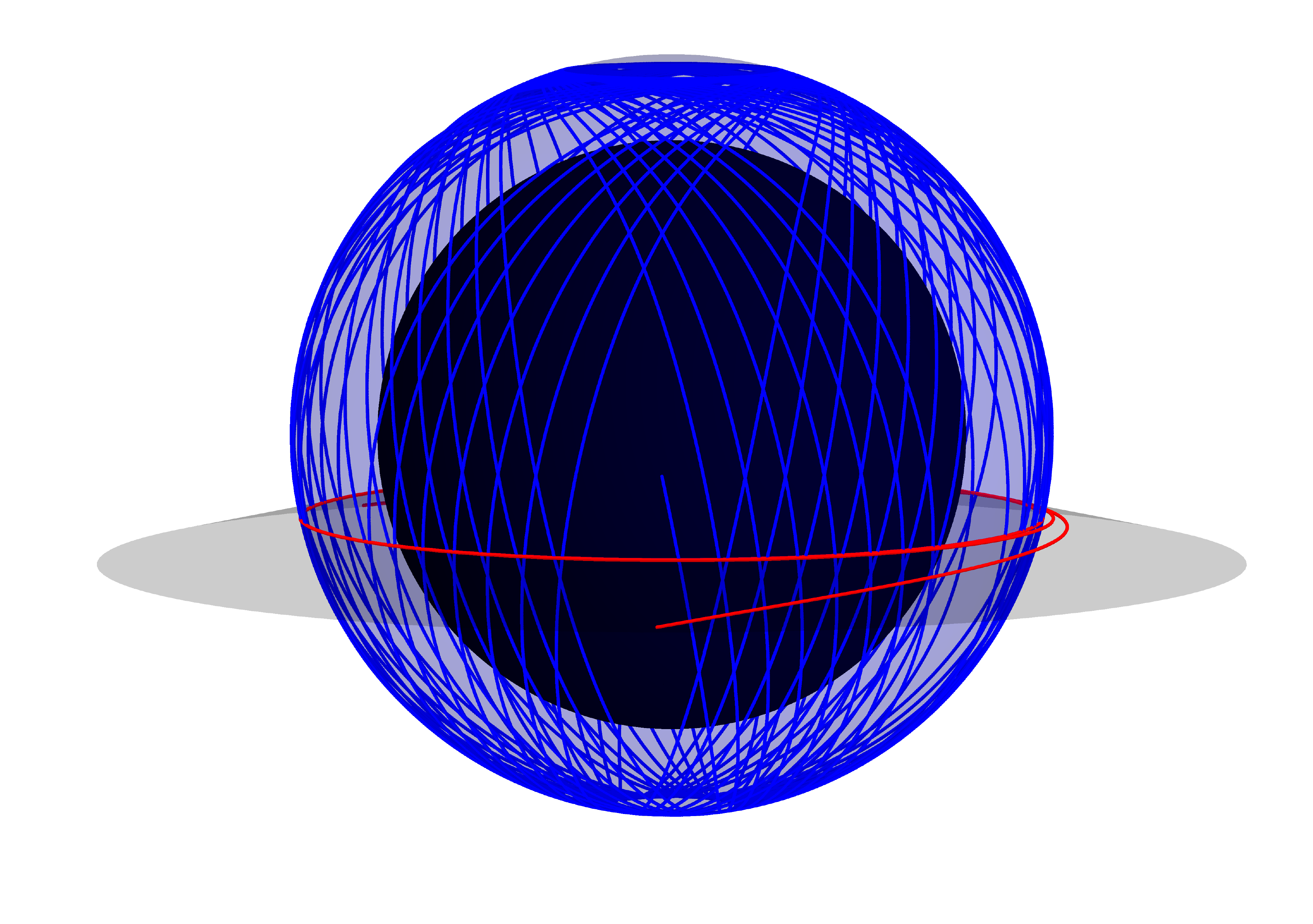}
\caption{Examples for two geodesics in the C-metric with acceleration parameter $\alpha=1/(4m)$. In the plot we use
standard spherical cordinates. The blue geodesic is a null geodesic on the photon sphere with $r_{i}=r_{ph}$,
$\vartheta_{i}=\pi/2$ and $\varphi_{i}=0$. The red geodesic is a geodesic on the photon cone with $r_{i}=3.9m$,
$\vartheta_{i}=\vartheta_{ph}$ and $\varphi_{i}=0$.}
\label{fig:photongeodesics}
\end{figure}

Determining $\varphi ( \lambda _f )$ is more awkward because the $\vartheta$ motion
may have arbitrarily many turning points. For determining the values of the Mino
parameters $\lambda _{\mathrm{turn,1}}$, $\lambda _{\mathrm{turn,2}}$ etc. that
lie in the interval between $\lambda _i =0$ and $\lambda _f$ we solve (\ref{eq:geocostheta})
for $\mathrm{d} \lambda /\mathrm{d} x$ and integrate from $x_i = \mathrm{cos} \, \vartheta _i$
up to the first turning point $x _{\mathrm{turn,1}}$. This is an elliptic integral of the
first kind that gives us $\lambda _{\mathrm{turn,1}}$. Integrating the same expression
from one turning point $x _{\mathrm{turn},k}$ up to the next $x _{\mathrm{turn},(k+1)}$
gives us the difference $\Delta \lambda =\lambda _{\mathrm{turn},(k+1)}
- \lambda _{\mathrm{turn},k}$. Note that this difference is independent of $k$. Then we
get $\varphi ( \lambda _f )$ from (\ref{eq:geophicosint}) where we have to break the integral
at each value $\lambda _{\mathrm{turn,1}}$, $\lambda _{\mathrm{turn,2}}=
\lambda _{\mathrm{turn,1}}+ \Delta \lambda$,  $\lambda _{\mathrm{turn,3}}=
\lambda _{\mathrm{turn,1}}+ 2 \Delta \lambda$ etc. that is smaller than $\lambda _f$
(for past-directed null geodesics: bigger than $\lambda _f$).

In figure \ref{fig:photongeodesics} we visualise two null geodesics. The blue geodesic
is a null geodesic on the photon sphere. We can see that it propagates back and forth between
the turning points of the $\vartheta$ motion while it precesses around the photon
sphere without crossing the axis. For special initial conditions the $\varphi$
motion and the $\vartheta$ motion have commensurable periods; then the geodesic
is closed. For generic initial conditions, the geodesic fills part of the photon
sphere densely. Note that the asymmetry with respect to the equatorial plane is clearly visible. --
The red geodesic is a null geodesic on the photon cone for which the $r$ coordinate
has a minimum $r_{\mathrm{min}}$ in close proximity to the photon sphere. If maximally
extended, this geodesic enters into the domain of outer communication over the
past acceleration horizon and leaves it over the future acceleration horizon.

\section{Gravitational lensing in the C-metric}\label{sec:lensing}
\subsection{Celestial coordinates}
Following the usual astronomical convention, we use latitude-longitude coordinates
on the celestial sphere of an observer. For this purpose we first fix a static observer
at the coordinates $(x_{O}^{\mu})=(t_{O}, r_{O}, \vartheta_{O}, \varphi_{O})$ between the black
hole horizon $r_{BH}$ and the acceleration horizon $r_{\alpha}$. Because of the
symmetry of the spacetime, it is irrelevant which values we choose for $t_O$ and
$\varphi _O$. At the position of the observer we now introduce an orthonormal tetrad
following Grenzebach et al. \cite{Grenzebach2015a,Grenzebach2016}:
\begin{eqnarray}
 e_{0}=\left.\frac{\Omega(r,\vartheta)}{\sqrt{Q(r)}}\partial_{t}\right|_{(x_{O}^{\mu})},
\end{eqnarray}
\begin{eqnarray}
 e_{1}=\left.\frac{\Omega(r,\vartheta)\sqrt{P(\vartheta)}}{r}\partial_{\vartheta}\right|_{(x_{O}^{\mu})},
\end{eqnarray}
\begin{eqnarray}
 e_{2}=-\left.\frac{\Omega(r,\vartheta)}{r\sin\vartheta\sqrt{P(\vartheta)}}\partial_{\varphi}\right|_{(x_{O}^{\mu})},
\end{eqnarray}
\begin{eqnarray}
 e_{3}=-\left.\Omega(r,\vartheta)\sqrt{Q(r)}\partial_{r}\right|_{(x_{O}^{\mu})}.
\end{eqnarray}
In general the tangent vector of a null geodesic in Mino parameterisation, $\eta ( \lambda)$, is
given by:
\begin{eqnarray}
 \frac{\mathrm{d}\eta}{\mathrm{d}\lambda}=\frac{\mathrm{d}t}{\mathrm{d}\lambda}\partial_{t}+\frac{\mathrm{d}r}{\mathrm{d}\lambda}\partial_{r}
 +\frac{\mathrm{d}\vartheta}{\mathrm{d}\lambda}\partial_{\vartheta}+\frac{\mathrm{d}\varphi}{\mathrm{d}\lambda}\partial_{\varphi} \, .
\end{eqnarray}
Introducing celestial coordinates $\Sigma$ (latitude) and $\Psi$ (longitude) at
the position of the observer the tangent vector can also be written as:
\begin{eqnarray}
 \frac{\mathrm{d}\eta}{\mathrm{d}\lambda}=\sigma\left(-e_{0}+\sin\Sigma\cos\Psi e_{1}+\sin\Sigma\sin\Psi e_{2}+\cos\Sigma e_{3}\right) \, .
\end{eqnarray}
The factor $\sigma$ is a normalisation constant given by:
\begin{eqnarray}
 \sigma=g\left(\frac{\mathrm{d}\eta}{\mathrm{d}\lambda},e_{0}\right) \, .
\end{eqnarray}
According to our convention of considering only lightlike geodesics with
$E>0$, the factor $\sigma$ has to be
negative. Note that the Mino parameter is defined only up to an affine
transformation. Therefore, we can assume without loss of generality that
$\sigma =-r_{O}^2/\Omega(r_{O},\vartheta_{O})^2$. Then comparing
the coefficients yields the constants of motion:
\begin{eqnarray}\label{eq:CoM}
\hspace*{-1.5cm}E=\frac{\sqrt{Q(r_{O})}}{\Omega(r_{O},\vartheta_{O})},~L_{z}=\frac{r_{O}\sqrt{P(\vartheta_{O})}\sin\vartheta_{O}\sin\Sigma\sin\Psi}{\Omega(r_{O},\vartheta_{O})},~K=\frac{r_{O}^2\sin^2\Sigma}{\Omega(r_{O},\vartheta_{O})^2}
\, .
\end{eqnarray}

\subsection{Angular radius of the shadow}
\begin{figure}[ht]
	\includegraphics[width=0.8\textwidth]{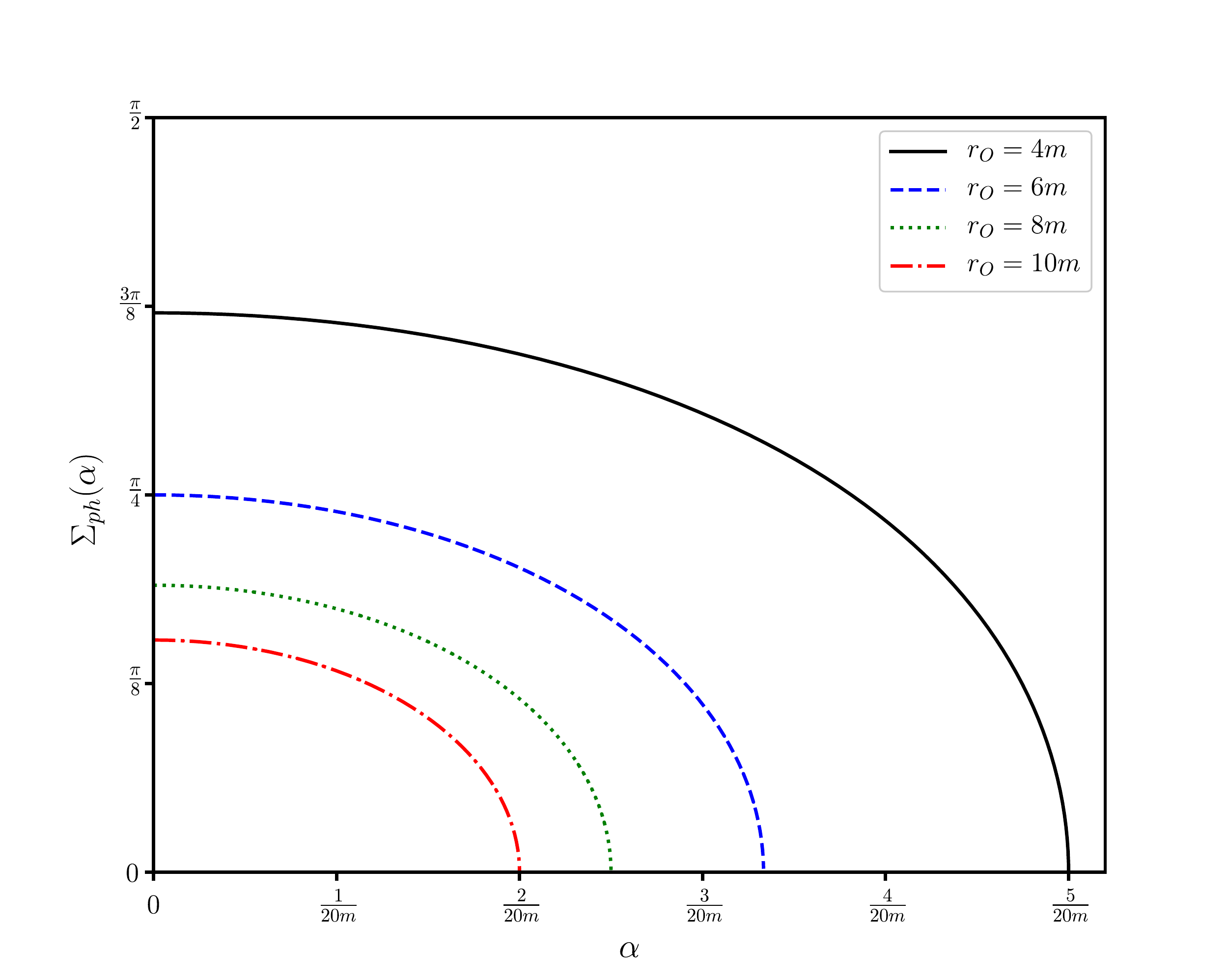}
\caption{Angular radius of the shadow for varying $\alpha$ for observers
located at $r_{O}=4m$ (solid black line), $r_{O}=6m$ (dashed blue line), $r_{O}=8m$
(dotted green line), and $r_{O}=10m$ (dash-dotted red line). Note that $\Sigma_{ph}$
does not depend on $\vartheta_{O}$.}
\label{fig:Photon}
\end{figure}
For the (idealised) definition of the shadow of a black hole we assume that there are
light sources everywhere in the universe but not between the observer and the black
hole. From the position of the observer, who is assumed to be in the domain of
outer communication, we follow all lightlike geodesics backwards in time. Note that
according to our conventions the Mino parameter \emph{decreases} along such
geodesics. There are two types of such geodesics: Those which go towards the
inner horizon and those which
go to the outer horizon. As the former will not meet a light source, according to our
assumption, we associate darkness with their initial directions.
By contrast, the latter will meet a light source, so we associate brightness with their initial
directions. Therefore, the boundary of the dark region, which is called the shadow,
is determined by light rays that go
to neither of the horizons. Each of these light rays is asymptotically spiralling towards a
lightlike geodesic that is contained in the photon sphere, so it must have the same
constants of motion as the limiting lightlike geodesic. For the limiting geodesic
(\ref{eq:geor}) has to hold with the left-hand side equal to zero and with $r= r_{ph}$
inserted on the right-hand side. On the other hand, the constants of motion of the
spiralling geodesic must satisfy (\ref{eq:CoM}). Combining these two observations
and solving for $\Sigma=\Sigma_{ph}$ gives the angular radius of the shadow:
\begin{eqnarray}\label{eq:Sigmaph}
\Sigma_{ph}=\arcsin\left(\frac{r_{ph}}{r_{O}}\sqrt{\frac{Q(r_{O})}{Q(r_{ph})}}\right)
\, .
\end{eqnarray}
This agrees with the result of Grenzebach et al. \cite{Grenzebach2015a} if the
latter is specified to the case of vanishing spin and vanishing NUT parameter.
Note that (\ref{eq:Sigmaph}) reduces to Synge's formula \cite{Synge1966} in the
Scharzschild limit ($\alpha\rightarrow 0$).

Here it is remarkable that we get an expression for $\Sigma=\Sigma_{ph}$ that
does not involve $\Psi$. Therefore, the shadow is circular. This could not have
been anticipated before the calculation was done because the C-metric is not
spherically symmetric. Similarly Grenzebach et al. \cite{Grenzebach2014,Grenzebach2015a}
discovered that the shadow of a rotating black hole with NUT charge or acceleration
is always symmetric to the equator on the observer's celestial sphere independent
of the latitude $\vartheta_{O}$ of the observer. This is also not intrinsically
obvious. While we do not have an explicit physical explanation for this fact, it
is very likely that in both cases it can be attributed to the existence of the
Carter constant $K$ for lightlike geodesics and thus to the associated hidden
symmetry.

In figure \ref{fig:Photon} we show $\Sigma_{ph}$ for different radius coordinates
$r_{O}$ of the observer.
As one would expect, the angular radius of the shadow decreases with increasing
$r_{O}$ and with increasing acceleration parameter. The latter observation
reflects the fact that, if $r_O$ is kept fixed, with increasing
$\alpha$ the acceleration horizon moves closer to the observer.

We have calculated the shadow for a static observer, i.e., we have assumed
that the 4-velocity of the observer is parallel to $\partial _t$. From this result
the shadow for a moving observer follows immediately by applying, on the tangent
space at each point, the special-relativistic aberration formula, cf. Grenzebach
\cite{Grenzebach2015c}. As the aberration formula maps circles onto circles, the
shadow is circular for \emph{every} observer.

In view of observations, it is unfortunate that in the C-metric the shadow is circular.
Therefore, we cannot distinguish the C-metric from the Schwarzschild spacetime,
or from any other spherically symmetric black-hole spacetime, by the \emph{shape}
of the shadow. It is true that the acceleration parameter has an effect on the
\emph{size} of the shadow. This, however, cannot be utilised as long as one
does not know the value of $r_O$ with high accuracy. Other lensing features by
which the C-metric could be distinguished from the Schwarzschild metric will be
discussed in the following subsections.

\subsection{The lens equation}
A lens equation, or lens map, for an arbitrary general-relativistic spacetime was
brought forward by Frittelli and Newman \cite{Frittelli1999}. For spherically
symmetric and static spacetimes, this lens equation was specified by
Perlick \cite{Perlick2004a}. In the following we apply the methodology
of the latter paper to the C-metric. As the C-metric is not spherically symmetric,
this requires some modifications.

\begin{figure}[ht]
\begin{center}
	\includegraphics[width=0.4 \textwidth]{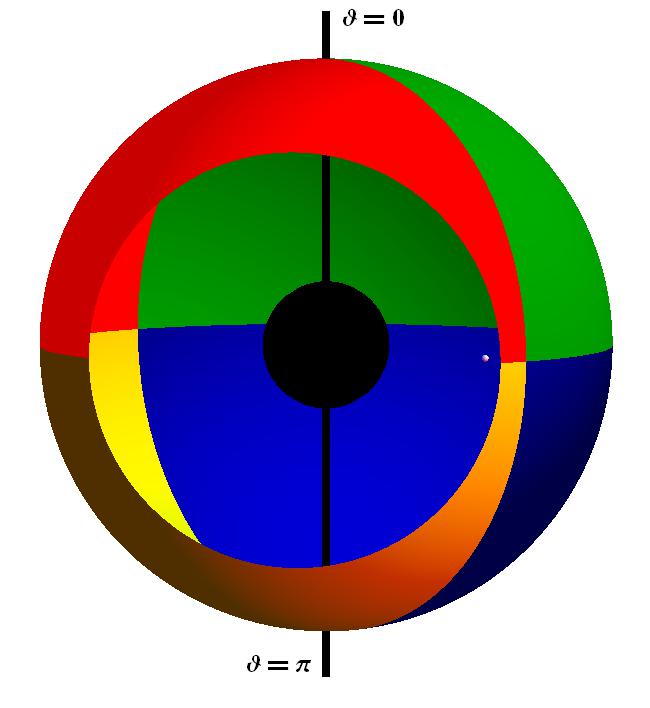}
\end{center}
\caption{Sphere with radius $r_L$ on which the light sources
are distributed. The black ball is the horizon, the vertical
black lines are the string and the strut, and the white dot
marks the position of the observer (here at $r_{O}=8m$,
$\vartheta_{O}=\pi/2$ and $\varphi_{O}=0$). What the observer
sees is the inner side of the sphere with radius $r_L$. The
colours represent the following coordinate ranges. $0\leq
\varphi_{L}<\pi$ and $0\leq\vartheta_{L}\leq\pi/2$: green;
$0\leq\varphi_{L}<\pi$ and $\pi/2<\vartheta_{L}\leq\pi$: blue;
$\pi\leq\varphi_{L}<2\pi$ and $0\leq\vartheta_{L}\leq\pi/2$:
red; $\pi\leq\varphi_{L}<2\pi$ and $\pi/2<\vartheta_{L}\leq\pi$:
yellow. The outer side of the sphere is shown darkened in the
picture because of lighting.}
\label{fig:sourcesphere}
\end{figure}

\begin{figure}[ht]
	\includegraphics[width=\textwidth]{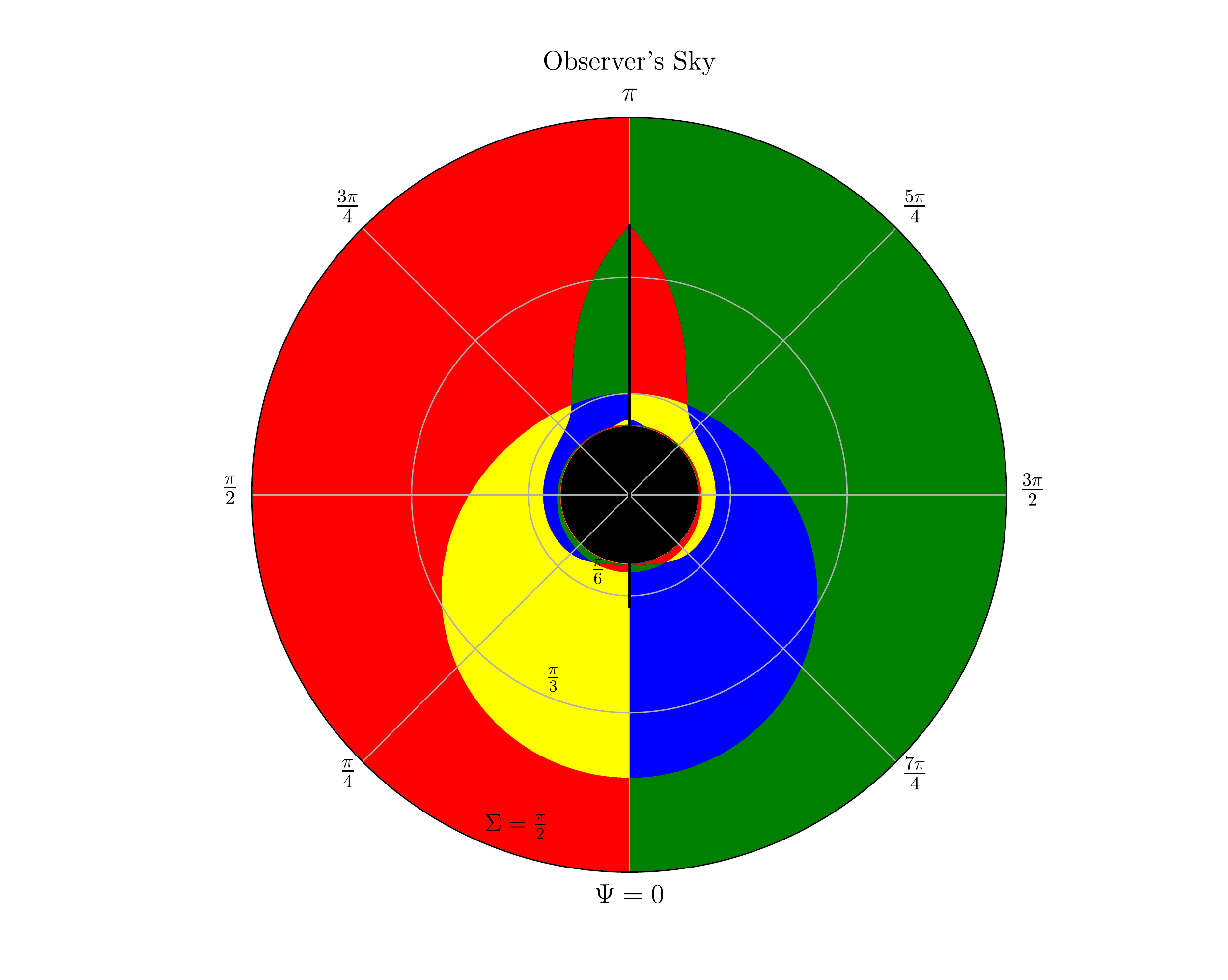}
\caption{Lens equation of light sources located on the two sphere $S_{L}^2$ with radius
coordinate $r_{L}=9m$ for an observer at $t_{O}=0$ located at $r_{O}=8m$,
$\vartheta_{O}=\pi/4$ and $\varphi_{O}=0$ in the C-metric with $\alpha=1/(10m)$.
The black lines at $\Psi=0$ and $\Psi=\pi$ indicate light rays crossing the axes at
least once.}
\label{fig:LensMapC2}
\end{figure}
\begin{figure}[ht]
	\includegraphics[width=\textwidth]{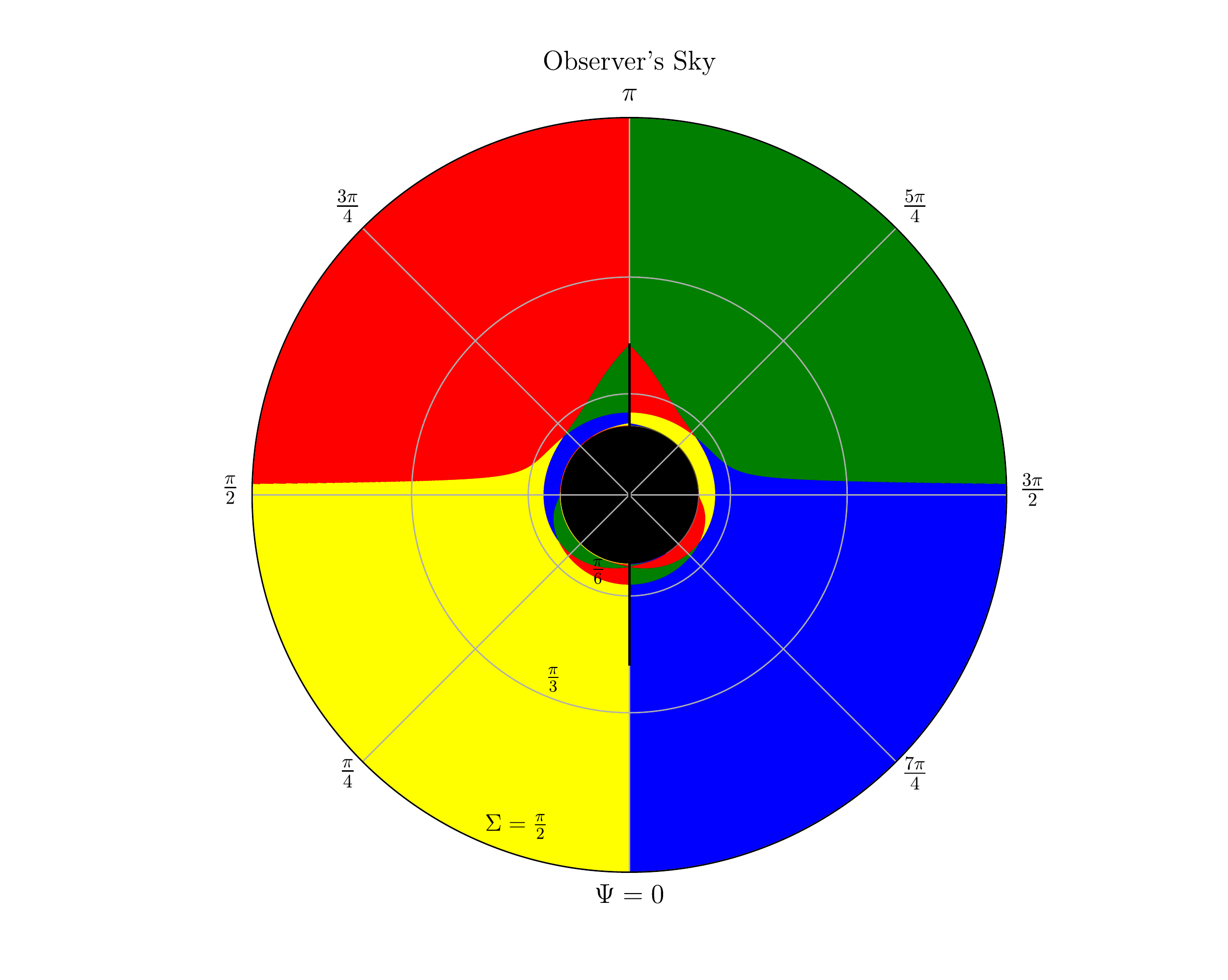}
\caption{Lens equation of light sources located on the two sphere $S_{L}^2$ with radius
coordinate $r_{L}=9m$ for an observer at $t_{O}=0$ located at $r_{O}=8m$,
$\vartheta_{O}=\pi/2$ and $\varphi_{O}=0$ in the C-metric with $\alpha=1/(10m)$.
The black lines at $\Psi=0$ and $\Psi=\pi$ indicate light rays crossing the axes at
least once.}
\label{fig:LensMapC3}
\end{figure}
\begin{figure}[ht]
	\includegraphics[width=\textwidth]{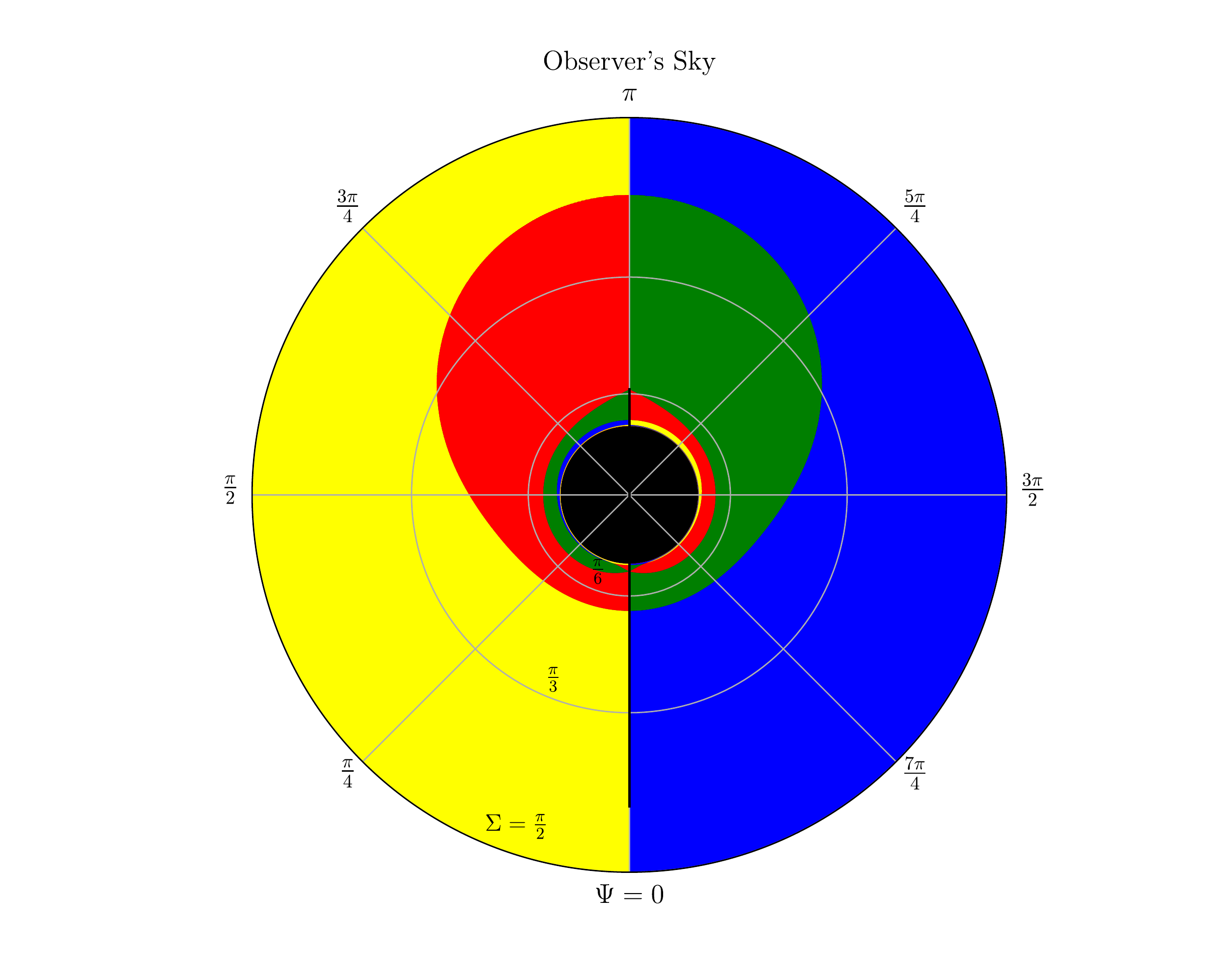}
\caption{Lens equation of light sources located on the two sphere $S_{L}^2$ with radius
coordinate $r_{L}=9m$ for an observer at $t_{O}=0$ located at $r_{O}=8m$,
$\vartheta_{O}=3\pi/4$ and $\varphi_{O}=0$ in the C-metric with $\alpha=1/(10m)$.
The black lines at $\Psi=0$ and $\Psi=\pi$ indicate light rays crossing the axes at
least once.}
\label{fig:LensMapC4}
\end{figure}

We first distribute static light sources on the two sphere $S^2_{L}$ with radius coordinate
$r_{L}>r_{O}$. Then we construct the past light cone for the static observer at the
coordinates $(x_{O}^{\mu})$. We follow all null geodesics from the observer
backwards in time until they intersect the two sphere $S^2_{L}$. Again, we
emphasise that according to our conventions the Mino parameter
\emph{decreases} along such geodesics. In general
not all such geodesics  intersect $S^2_{L}$ but for those which do we
can construct the lens equation as defining a map from the angular coordinates
$\Sigma$ and $\Psi$ on the observer's celestial sphere to the angular coordinates
$\vartheta_{L}(\Sigma,\Psi)$ and $\varphi_{L}(\Sigma,\Psi)$ on the sphere of the
light sources:
\begin{eqnarray}
(\Sigma,\Psi)\rightarrow(\vartheta_{L}(\Sigma,\Psi),\varphi_{L}(\Sigma,\Psi)) \, .
\end{eqnarray}
So the lens map is a map from part of a 2-sphere to a 2-sphere. In contrast to the
spherically symmetric case which was treated by Perlick \cite{Perlick2004a}, this map
is not rotationally symmetric, i.e., it does not reduce to a map from one angle to another
angle.

We calculate $\vartheta_{L}(\Sigma,\Psi)$ and $\varphi_{L}(\Sigma,\Psi)$ from the
analytic solutions of the equations of motion presented in the last section by
setting $(x_{i}^{\mu})=(x_{O}^{\mu})$. Note that in this way we get a fully
analytic lens map. The only unknown we have to eliminate is the Mino parameter
$\lambda _L$ of the geodesic when it meets the sphere at $r_L$. (Without loss of
generality, we assume that all geodesics start at $\lambda _O =0$ from the
observer's position and end at $\lambda_{L}<\lambda _O$ at the position of the
light source.) Since we fixed the radius coordinates $r_{O}$ and $r_{L}$ of the
observer and of the light sources, respectively, we can use them to calculate
$\lambda _L$. For this purpose we first express $E$ and $K$ in (\ref{eq:geor})
by (\ref{eq:CoM}). Then we integrate and obtain $\lambda_L$. Here we have to
distinguish the case with a turning point from the case without turning point.
We will assume that $r_O$ and $r_L$ are both bigger than $r_{ph}$. Then a turning
point cannot be a maximum. For null geodesics with a minimum as a turning point
$\lambda_L$ reads
\begin{eqnarray}
\lambda_L=
\left(\int_{r_{O}}^{r_{min}}-\int_{r_{min}}^{r_{L}}\right)
\frac{\Omega(r_{O},\vartheta_{O})\mathrm{d}r'}{\sqrt{Q(r_{O})
r'^4-r_{O}^2\sin^2\Sigma r'^2Q(r')}},
\end{eqnarray}
while for null geodesics without turning points it reads:
\begin{eqnarray}
\lambda_L=-\int_{r_{O}}^{r_{L}}\frac{\Omega(r_{O},\vartheta_{O})\mathrm{d}r'}{\sqrt{Q(r_{O})r'^4-r_{O}^2\sin^2\Sigma r'^2Q(r')}}.
\end{eqnarray}
As before we can rewrite the integral on the right-hand side (which, for some
special cases, reduces to an elementary integral) as an elliptic integral of the
first kind. After having calculated the Mino parameter we then proceed as
described in section \ref{subsec:examples} to calculate $\vartheta_{L}(\Sigma,\Psi)$
and $\varphi_{L}(\Sigma,\Psi)$.

Figures \ref{fig:LensMapC2}-\ref{fig:LensMapC4} show the lens map for different
observer positions in the C-metric with $\alpha = 1/(10 m)$.  The celestial sphere
of the observer is represented in stereographic projection, with the direction towards the
black hole at the centre. We have chosen the values $r_{O}=8m$, $r_{L}=9m$,
$\vartheta_{O}=\pi/4$ (figure \ref{fig:LensMapC2}), $\vartheta_{O}=\pi/2$
(figure \ref{fig:LensMapC3}) and $\vartheta_{O}=3\pi/4$  (figure \ref{fig:LensMapC4}).
The sphere at $r_L$ is divided into four quarter-spheres which are painted red, green,
blue and yellow, respectively, see figure \ref{fig:sourcesphere}. This colouring follows
the convention in Bohn et
al. \cite{Bohn2015}. In the following we refer to the area on the sky where
$\pi/2 < \Psi <3\pi/2$ as the northern hemisphere and to the rest of the sky
as the southern hemisphere.

Figure \ref{fig:LensMapC2} shows the lens equation for $\vartheta_{O}=\pi/4$.
There are, for each light source, in principle infinitely many images. We say that an
image is of order $k$ if the $\varphi$ coordinate of the corresponding geodesic
ranges over an interval $\Delta \varphi$ with $(k-1) \pi < | \Delta \varphi | < k \pi$.
In the centre of the figure we see the shadow. In the outermost part, coloured in red/yellow
on the left and in green/blue on the right, there are the primary images ($k=1$). When
we move closer towards the shadow this order reverses and thus these are secondary
images, $k=2$. Moving further in we can also see images of order 3 and 4, where
the latter are only visible when zooming into the online version of the figure. The borderlines
between images of different orders mark the critical curves. In the Schwarzschild
spacetime these are circles (``Einstein rings''). In the C-metric they have a rather complicated
shape that varies for the transitions between images of different orders.

We also see that on the northern hemisphere images of second and third order can
be observed much further away from the shadow than on the southern hemisphere. In
particular around $\psi=0$ higher-order images are visible only very close to the
shadow. We also see that light rays travelling towards $\vartheta=0$ cover the
same interval of $\Delta\varphi$ faster than light rays travelling towards $\vartheta=\pi$.

In figure \ref{fig:LensMapC3} we see the lens map for $\vartheta_{O}=\pi/2$.
This plot clearly demonstrates that in the C-metric there is no symmetry with
respect to the plane $\vartheta = \pi /2$. Correspondingly, past-oriented light rays
that leave the observer with $| \mathrm{sin} \, \Psi | =1$ do \emph{not}
necessarily meet the sphere of light sources at $\vartheta _L = \pi /2$. In
addition we see that, compared to figure \ref{fig:LensMapC2},
on the northern hemisphere most features, in particular images of second order or
higher, are more centred around the shadow while around $\Psi=0$ they now already
appear at a larger distance to the shadow.

In figure \ref{fig:LensMapC4} we see the lens map for $\vartheta_{O}=3\pi/4$.
Since the observer is now located at $\vartheta_{O}>\pi/2$, on the outer part of
the image the colours red/yellow and green/blue are interchanged with respect
to figure \ref{fig:LensMapC2}. On the northern hemisphere images beyond second order can
now be observed at angles even closer to the shadow while on the southern hemisphere
the images move further away from the shadow.

The broken symmetry with respect to the equatorial plane distinguishes the
C-metric from the Schwarzschild metric and from all other spherically symmetric
black-hole metrics. In particular, in the C-metric multiple images of a light source are not
located on a great circle through the centre of the shadow.   This is an observable
feature that could become relevant if and when multiple images produced by a
black hole are detected.

\subsection{Redshift}
\begin{figure}[ht]
	\includegraphics[width=\textwidth]{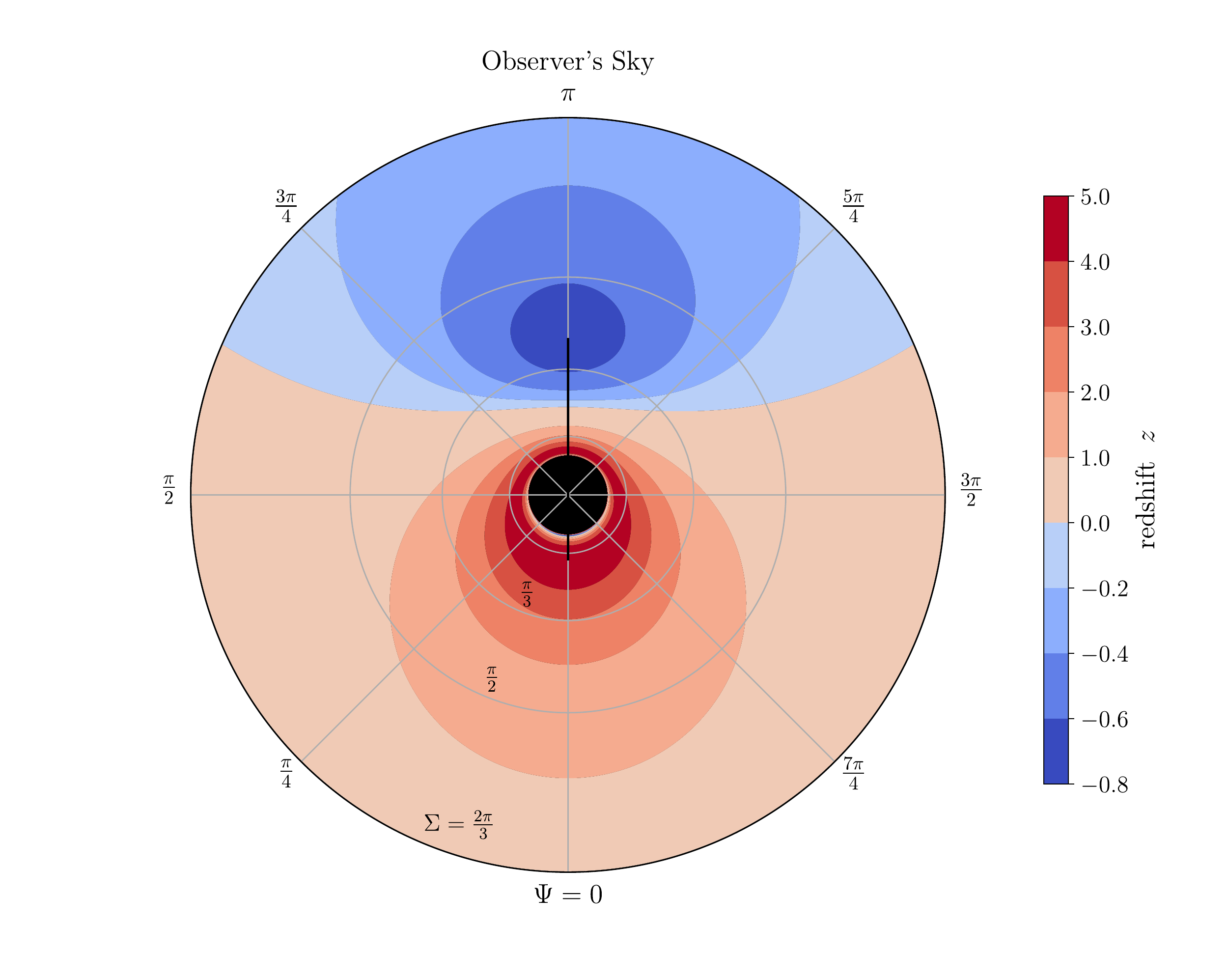}
\caption{Redshift of light sources located on the two sphere $S_{L}^2$ with radius
coordinate $r_{L}=9m$ measured by an observer at $r_{O}=8m$,
$\vartheta_{O}=\pi/4$ in the C-metric with $\alpha=1/(10m)$. The
black lines at $\Psi=0$ and $\Psi=\pi$ indicate light rays crossing the axes at
least once.}
\label{fig:Redshift2}
\end{figure}
\begin{figure}[ht]
	\includegraphics[width=\textwidth]{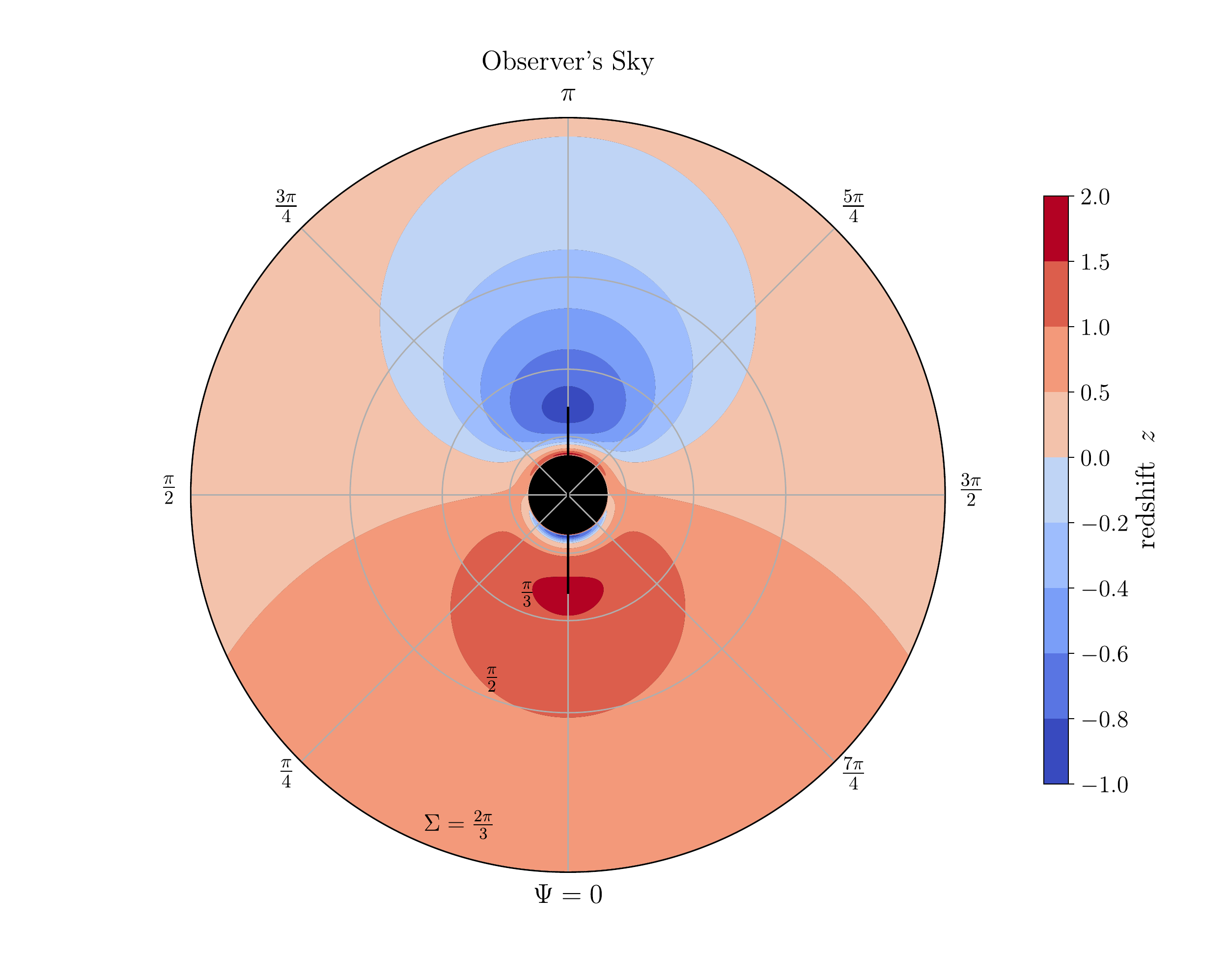}
\caption{Redshift of light sources located on the two sphere $S_{L}^2$ with radius
coordinate $r_{L}=9m$ measured by an observer at $r_{O}=8m$,
$\vartheta_{O}=\pi/2$ in the C-metric with $\alpha=1/(10m)$. The
black lines at $\Psi=0$ and $\Psi=\pi$ indicate light rays crossing the axes at
least once.}
\label{fig:Redshift3}
\end{figure}
\begin{figure}[ht]
	\includegraphics[width=\textwidth]{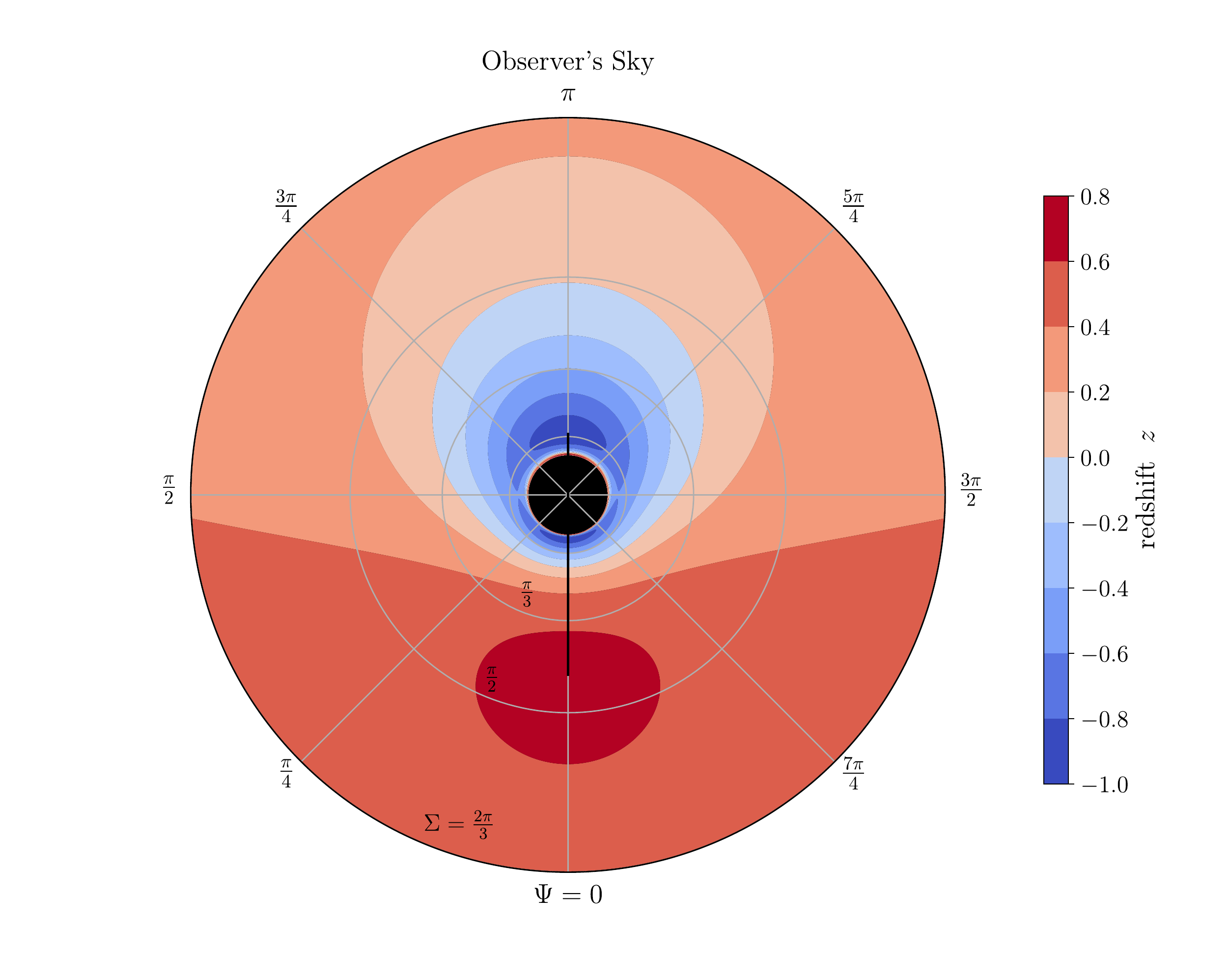}
\caption{Redshift of light sources located on the two sphere $S_{L}^2$ with radius
coordinate $r_{L}=9m$ measured by an observer at $r_{O}=8m$,
$\vartheta_{O}=3\pi/4$ in the C-metric with $\alpha=1/(10m)$. The
black lines at $\Psi=0$ and $\Psi=\pi$ indicate light rays crossing the axes at
least once.}
\label{fig:Redshift4}
\end{figure}
The redshift is one of the few parameters that are directly accessible to observations.
If we can identify emission lines in the spectrum of a light source, we can use them to directly
determine the redshift. When we are able to determine the redshift for sufficiently many
pixels in an astrophysical image we can construct a redshift map and draw
conclusions on the underlying spacetime by comparing with theoretical predictions.

In the C-metric we consider again, as for the lens map, a static observer at coordinates
$r_O$ and $\vartheta _O$ and a static light source at coordinates $r_L$ and
$\vartheta _L (\Sigma, \Psi )$. The redshift for this situation can be found, e.g.,
in the book of Straumann \cite{Straumann2013}, pp.~45, and reads
\begin{eqnarray}
z=\sqrt{\frac{\left.g_{tt}\right|_{x_{O}}}{\left.g_{tt}\right|_{x_{L}}}}-1.
\end{eqnarray}
When inserting the $g_{tt}$ of the C-metric this equation becomes:
\begin{eqnarray} \label{eq:redshift}
z=\sqrt{\frac{Q(r_{O})}{Q(r_{L})}}\frac{\Omega(r_{L},\vartheta_{L}(\Sigma,\Psi))}{\Omega(r_{O},\vartheta_{O})}-1.
\end{eqnarray}
We can immediately see that, unlike in spherically symmetric and static spacetimes,
in the C-metric the redshift does not only depend on the radius coordinates $r_{O}$
and $r_{L}$ but also on $\vartheta_{O}$ and $\vartheta_{L}(\Sigma,\Psi)$ (and via
the latter also on the celestial coordinates). Therefore, with $r_O$, $\vartheta _O$ and
$r_L$ fixed, the redshift can be considered as a function on the observer's celestial sphere.

Figures \ref{fig:Redshift2}-\ref{fig:Redshift4} show the redshift $z$ on the observer's
celestial sphere for the same situations as in figures \ref{fig:LensMapC2}-\ref{fig:LensMapC4}.
Again, we consider the C-metric with $\alpha = 1/(10m)$. For the chosen numbers of
$r_O = 8m$ and $r_L = 9m$ the reference redshift (or better
blueshift) for the Schwarzschild metric ($\alpha=0$) is $z=-0.018$.

We can see that in the C-metric the redshift factor varies in the shown images
between $z=-1$ and $z=5$. (From (\ref{eq:redshift}) it is clear that $z$ is always
bigger than -1.) As for the lens map we can clearly
see the axisymmetry of the redshift map. However, the most salient feature is
that in contrast to the Schwarzschild metric, where we always observe a blueshift
if $r_O < r_L$, for the C-metric we observe both redshifts and blueshifts.

Figure \ref{fig:Redshift2} shows the redshift map for an observer at $\vartheta_{O}=\pi/4$.
We can see that the northern hemisphere is dominated by blueshifts and the southern
hemisphere is dominated by redshifts. In addition we can recognise a small and narrow
region of blueshifts close to the shadow on the southern hemisphere. (If we zoom
in we see a third region of blueshifts even closer to the shadow on the northern
hemisphere.) Note that in the redshift equation (\ref{eq:redshift})
the root is always constant, so the variation of the redshift factor comes purely
from the conformal factor $\Omega(r,\vartheta)$. For $\vartheta_{O}=\pi/4$
we have $\Omega(r_{O},\vartheta_{O})<1$. In addition we have
$1/10\leq\Omega(r_{L},\vartheta_{L})\leq19/10$
and in particular $\Omega(r_{L},\vartheta_{L})<\Omega(r_{O},\vartheta_{O})$ for
$\vartheta_{L}\leq\vartheta_{O}$.
Thus the closer the light source to the axis $\vartheta=0$ the higher the blueshift
on the northern hemisphere and the closer the light source to the axis $\vartheta=\pi$
the higher the redshift on the southern hemisphere.

Figure \ref{fig:Redshift3} shows the redshift map for $\vartheta_{O}=\pi/2$. As
for the lens map we clearly see the asymmetry with respect to the line $| \mathrm{sin} \, \Psi | = 1 $.
The range covered by $z$ decreases while the maximum blueshift increases. In
addition we observe that on the northern hemisphere the region of blueshift shrinks
while it becomes slightly larger on the southern hemisphere. The reduction of the
range can be easily understood since in this case we have $\Omega(r_{O},\vartheta_{O})=1$
which is thus larger than for the previous figure.

Figure \ref{fig:Redshift4} shows the redshift map for $\vartheta_{O}=3\pi/4$. The
range of the redshift factor is further reduced since $\Omega(r_{O},\vartheta_{O})>1$.
The outer part of the figure is now dominated by redshifts. Regions with strong blueshifts
now only occur around the shadow and the formerly separated regions in the northern
hemisphere and the southern hemisphere connect. Within the region of blueshifts
we can still see a small crest-like region of redshifts but it is now much smaller
than in the previous figures.

Considering static light sources on a sphere with radius $r_L$ is a convenient
way of illustrating the characteristic lensing features in a spacetime. However,
light sources which we can actually observe around a black hole are, of
course, not located on such a sphere. Therefore, for comparison with
observations one has to consider more realistic light sources, e.g. radiating matter
that falls towards the black hole or rotates in an accretion disc. It is very
well possible to provide redshift maps for such situations in the C-metric but
we will not do this here.

\subsection{Travel time}
The travel time measures, in terms of the time coordinate $t$, how long a
light ray needs to travel along a null geodesic from the light source to the observer.
For a light ray emitted by a light
source at the time coordinate $t_{L}$ and detected by an observer at the time
coordinate $t_{O}$ it reads
\begin{eqnarray}
T=t_{O}-t_{L}.
\end{eqnarray}
If an observer sees two images of a light source whose emission
varies temporally,  then the travel time \emph{difference}, known as the
\emph{time delay}, is directly observable. Of course, one has to convert coordinate time
into proper time of the observer because only the latter is measurable.

For our specific purposes we set $t_{O}=0$. After replacing the integration limits in
(\ref{eq:geotint}) and substituting for $E$ and $K$ from (\ref{eq:CoM}) we obtain
the general form of the travel time integral:
\begin{eqnarray}
T(\Sigma)=\int_{r_{O}...}^{...r_{L}}\frac{\sqrt{Q(r_{O})}r'^2\mathrm{d}r'}{Q(r')\sqrt{Q(r_{O})r'^4-r_{O}^2\sin^2(\Sigma) r'^2 Q(r')}}.
\end{eqnarray}
It is described in section \ref{subsec:time} how this travel time integral is to be
adapted to each of the different types of $r$ motion.

We see that, if $r_O$ and $r_L$ are fixed, the travel time depends only on the celestial
latitude $\Sigma$ and not on the celestial longitude $\Psi$. Again, this result is non-trivial
because the spacetime is only axisymmetric but not spherically symmetric.
\begin{figure}[ht]
	\includegraphics[width=0.8\textwidth]{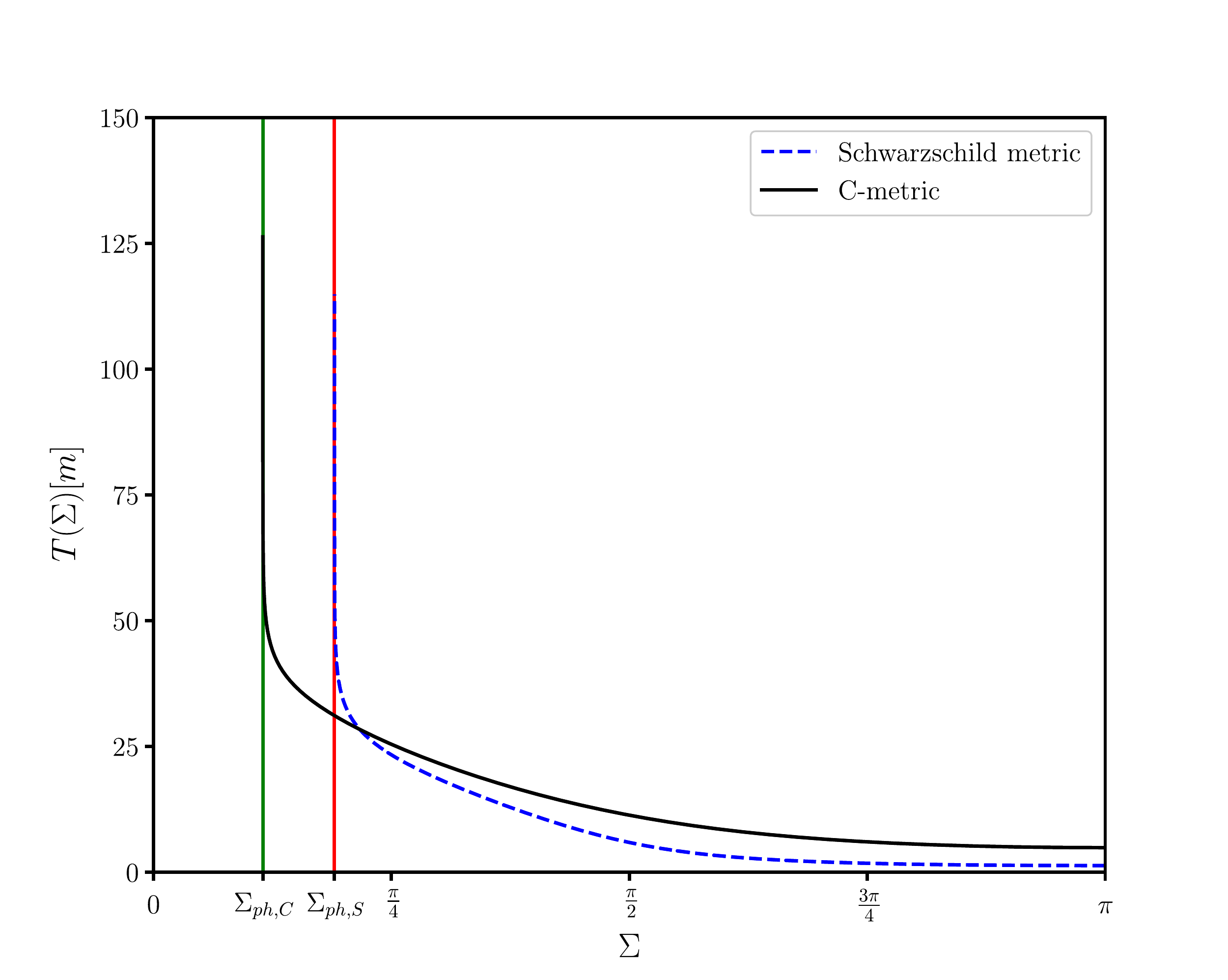}
\caption{Travel time for a light ray emitted by a light source at $r_{L}=9m$
and detected by an observer at $r_{O}=8m$ in the Schwarzschild metric (dashed blue line)
and in the C-metric with $\alpha=1/(10m)$ (solid black line). The red and green lines
mark the angular radius of the shadow in the Schwarzschild metric $\Sigma_{ph,S}$ and
in the C-metric $\Sigma_{ph,C}$, respectively.}
\label{fig:TravelTime}
\end{figure}
Figure \ref{fig:TravelTime} shows the travel time for a light ray emitted by a
light source located at $r_{L}=9m$ and detected by an observer located at $r_{O}=8m$
in the C-metric with $\alpha=1/(10m)$ and, for the sake of comparison, in the Schwarzschild
metric ($\alpha=0$). In either case the travel time diverges if $\Sigma$ approaches the
angular radius of the shadow. The latter is smaller in the C-metric than in the
Schwarzschild metric. However, except for $\Sigma$ close to $\Sigma_{ph,C}$
the travel time of a light ray in the C-metric is longer than for a light ray observed under
the same angle $\Sigma$ in the Schwarzschild metric. Thus one consequence of a non-zero
acceleration parameter is an increase of travel time. In addition between $\Sigma=\pi/4$ and
$\Sigma=\pi/2$ the travel time increases more slowly for the C-metric than for the
Schwarzschild metric.

We have already mentioned that only time delays, i.e., travel time \emph{differences},
in a multiple-imaging situation are observable.
This is routinely done for quasars that are multiply imaged by galaxies, see e.g.
\cite{Koptelova2012,Fohlmeister2013}. Unfortunately, light sources
multiply-imaged by black holes have not yet been observed, so we have to
wait for future generations of telescopes with higher resolutions before we
can confront the results of this section with observations.

\section{Conclusions}
In the first part of the paper we discussed and solved the equations of motion for lightlike
geodesics in the domain of outer communication of the C-metric, using Jacobi's elliptic
functions and the elliptic integrals of first and third kind. Whereas in this work we
derived the solutions in Boyer-Lindquist-like coordinates, this approach can
also be applied to other coordinate systems as, e.g., done very recently in Lim \cite{Lim2021}
for the C-metric with a cosmological constant. In particular we derived
and discussed the properties of the photon sphere and of the photon cone. It is
true that the results on the photon sphere and on the photon cone are not entirely
new: The photon sphere was discussed by Grenzebach et
al. \cite{Grenzebach2015a,Grenzebach2016} within a bigger class of spacetimes
and by Gibbons and Warnick \cite{Gibbons2016} from a more geometrical
point of view; some aspects of the photon cone were presented by
Alrais Alawadi et al. \cite{AlraisAlawadi2021}. However, we believe
that the approach presented here, paralleling the photon sphere and the photon cone
by using the potentials $V_r$ and $V_{\vartheta}$, has some merits and facilitates
understanding of the important features.

Based on our representation of the lightlike geodesics in terms of elliptic integrals
and elliptic functions, we then gave a detailed account of the lensing features in
the C-metric, for static observers and static light sources in the domain of outer
communication. More precisely, we calculated the shadow, we worked out a lens
equation and we discussed the redshift and the travel time. For achieving
these results it was crucial to introduce latitude-longitude coordinates on the oberver's
celestial sphere which are adapted to an orthonormal tetrad; in this respect our
methodology closely followed Grenzebach et al. \cite{Grenzebach2015a} .

In particular, we found that the shadow is circular, for \emph{every} observer
in the domain of outer communication. This result is highly non-trivial
because the C-metric is not spherically symmetric.  We found that the
angular radius of the shadow shrinks with increasing acceleration parameter
if we keep the radius coordinate $r_O$ of the observer fixed. This result is,
however, of limiting usefulness in view of observations as long as we do not
know $r_O$ (in units of $m$) with high accuracy.

As a main result of this paper we derived a lens equation for the C-metric and
plotted it on the observer's celestial sphere. We emphasise that our illustrations
of this lens equation are based on the analytical solutions of the geodesic equation
in terms of elliptic functions and \emph{not} on ray tracing. When zooming into
the online version of these illustrations one can identify images up to fourth order.
The images also clearly show the critical curves which are the borderlines
between images of different orders. In spherical symmetric spacetimes the
critical curves are circles (``Einstein rings'') but in the C-metric they have a
more complicated shape.

For an observer at $r_O$ and light sources at $r_L$, we calculated and plotted
the redshift $z$ as a function on the observer's celestial sphere. In the Schwarschild
metric this function is a constant. (For $r_O < r_L$ it is actually a blueshift.)
In the C-metric, however, it is a function of both celestial coordinates, latitude
$\Sigma$ and longitude $\Psi$. The range of $z$ depends, of course, on
the numerical value of the acceleration parameter $\alpha$ and goes to zero
for $\alpha \to 0$. The lowest values of $z$ (in our case blueshifts) occur near
the string at $\vartheta=0$ and the highest values of $z$ (in our case redshifts)
occur near the strut at $\vartheta=\pi$.

Lastly we derived the travel time. As in the Schwarzschild spacetime,
for an observer at $r_O$ and light sources at $r_L$ the travel time depends only
on the celestial latitude $\Sigma$. We observed that, keeping $r_O$, $r_L$ and
$\Sigma$ fixed, the travel time is bigger in the C-metric than in the Schwarzschild
metric unless $\Sigma$ is close to the shadow radius.

In this paper we wanted to provide some theoretical background for
distinguishing accelerating black holes from  non-accelerating ones
by way of lensing observations. It is unfortunate that the \emph{shape}
of the shadow is not such a distinguishing feature because the
C-metric predicts a circular shadow, just as the Schwarzschild metric.
The \emph{size} of the shadow is not a particularly useful observable
because it depends not only on the spacetime parameters but also
on $r_O$ which is not known with very high accuracy. If we ever
see multiple images produced by a black hole, we will be able to
determine the parameters of the black hole with much higher
accuracy than now: Combining measurements of the redshifts, of
the time delay and of the positions in the sky of the images might
give us a chance to distinguish accelerating black holes from
non-accelerating ones.

\section*{Acknowledgment}
We acknowledge support from Deutsche Forschungsgemeinschaft
within the Research Training Group 1620 Models of Gravity.
We also would like to express our gratitude to all contributors
to the Julia project and in particular the authors of the
packages Elliptic, Blosc, HDF5 and PyPlot. We also would like to
thank F. Willenborg for his help with fixing a bug with the layering
in the redshift and lens map images.

\appendix
\section{Integrals for geodesic motion}
In this appendix we briefly demonstrate how to calculate the elementary integrals
associated with the radius coordinate $r$ and the time coordinate $t$ for lightlike
geodesics asymptotically coming from (or going to) the photon sphere and the elliptic
integrals associated with $t$.

\subsection{Elementary integrals}
While calculating the solutions for $r(\lambda)$ and $t(\lambda)$ for lightlike
geodesics asymptotically coming from (or going to) the photon sphere in Sections
3.2.3 and 3.5.2 we encountered elementary integrals that can be easily solved using
coordinate transformations. While in each case the parameters of the integrals
vary we can generally transform them into one of the following two forms:
\begin{eqnarray}\label{eq:int}
I_{1}=\int\frac{\mathrm{d}y}{(y-a)\sqrt{y-y_{1}}},~~~I_{2}=\int\frac{\mathrm{d}y}{(a-y)\sqrt{y-y_{1}}} \, .
\end{eqnarray}
Recall that $y_{1}$ is related to the negative root $r_{4}$ via
(\ref{eq:rsub1}). Here we have $a=y_{ph}$ in (\ref{eq:geoyint}) and $I_{ph,4\pm}$
in (\ref{eq:geotsolphoton2}) and $a=\frac{K}{6}$, $a=-\frac{1-6\alpha m}{12}K$ and
$a=-\frac{1+6\alpha m}{12}K$ in $I_{ph,1}$, $I_{ph,2}$ and  $I_{ph,3}$ in (\ref{eq:geotsolphoton2}),
respectively. These integrals can now be solved using simple transformations all
containing the constant parameter $a-y_{1}>0$. For $y>a$ we now substitute $z=y-a$
and obtain for $I_{1}$:
\begin{eqnarray}\label{eq:intsol1}
I_{1}=\int\frac{\mathrm{d}z}{z\sqrt{z+a-y_{1}}}=-\frac{2}{\sqrt{a-y_{1}}}\mathrm{arcoth}\left(\sqrt{\frac{y-y_{1}}{a-y_{1}}}\right) \, .
\end{eqnarray}
Analogously for $y<a$ we substitute $z=y-y_{1}$ and obtain for $I_{2}$:
\begin{eqnarray}\label{eq:intsol2}
I_{2}=\int\frac{\mathrm{d}z}{(a-y_{1}-z)\sqrt{z}}=\frac{2}{\sqrt{a-y_{1}}}\mathrm{artanh}\left(\sqrt{\frac{y-y_{1}}{a-y_{1}}}\right) \, .
\end{eqnarray}
Now we solve (\ref{eq:rsub1}) for $y$ and replace all terms containing $y$,
$y_{ph}$ or $y_{1}$ in (\ref{eq:intsol1})-(\ref{eq:intsol2}) to finally obtain the integral
on the right-hand side of (\ref{eq:geoyint}) and $I_{ph,1}-I_{ph,4\pm}$ in (\ref{eq:geotsolphoton2}).

\subsection{Elliptic integrals}
An elementary introduction to elliptic integrals and functions can be found in
the book of Hancock \cite{Hancock1917}. In this paper we only need the incomplete
elliptic integrals of first and third kind. In Legendre form these integrals read
(in order):
\begin{eqnarray}\label{eq:EllipticIntegrals}
\hspace*{-2.25cm}F_{L}(\chi,k)=\int_{0}^{\chi}\frac{\mathrm{d}\chi'}{\sqrt{1-k\sin^2\chi'}},~~\Pi_{L}(\chi,k,n)=\int_{0}^{\chi}\frac{\mathrm{d}\chi'}{\left(1-n\sin^2\chi'\right)\sqrt{1-k\sin^2\chi'}} \, .
\end{eqnarray}
Here, $0<k<1$ is the square of the elliptic modulus, $n\in \mathbb{R}$ is an
additional parameter and $\chi$ is called the amplitude of the elliptic integral.

In addition to the elementary integrals of the previous section in (\ref{eq:geotint}),
we also encountered one elliptic integral not directly adopting the Legendre form
after coordinate transformation (\ref{eq:georsub4}). In its original form this
integral reads:
\begin{eqnarray}\label{eq:ellipticJ}
J(\chi_{i},\chi,k,n)=\int_{\chi_{i}}^{\chi}\frac{\mathrm{d}\chi'}{(1+n\cos\chi')\sqrt{1-k\sin^2\chi'}}
\, .
\end{eqnarray}
Now we outline how this integral can be rewritten in terms of the elliptic
integrals of first and third kind (\ref{eq:EllipticIntegrals}). For the calculations
we suppress the limits to ease the calculations and to reduce the length of the
obtained results.

As a first step we expand by $1-n\cos\chi$ and split the result into a term containing
the elliptic integral of third kind and a second term $L(\chi,k,n)$ only containing
an elementary integral:
\begin{eqnarray}\label{eq:ellipticJint}
J(\chi,k,n) & =
\int\frac{\mathrm{d}\chi'}{(1+n\cos\chi')\sqrt{1-k\sin^2\chi'}}
\\
& =\frac{1}{1-n^2}\left(\Pi_{L}\left(\chi,k,\frac{n^2}{n^2-1}\right)-nL(\chi,k,n)\right) \, . \nonumber
\end{eqnarray}
The computation of $L(\chi,k,n)$ requires several case-by-case analyses and is
rather lengthy. Thus we do not reproduce it here and only provide the final result:
\begin{eqnarray} \label{eq:ellipticL}
\hspace*{-1.5cm}L(\chi,k,n)&=\int\frac{\cos\chi'\mathrm{d}\chi'}{\left(1-\frac{n^2}{n^2-1}\sin^2\chi'\right)\sqrt{1-k\sin^2\chi'}}\\
&=\frac{1}{2}\sqrt{\frac{n^2-1}{n^2(1-k)+k}}\ln\left(\frac{\sqrt{\frac{n^2(1-k)+k}{n^2-1}}\sin\chi+\sqrt{1-k\sin^2\chi}}{\left|\sqrt{\frac{n^2(1-k)+k}{n^2-1}}\sin\chi-\sqrt{1-k\sin^2\chi}\right|}\right) \, . \nonumber
\end{eqnarray}
In our case we always integrate over the black hole horizon $r_{BH}$.
Here, $\Pi_{L}(\chi,k,n^2/(n^2-1))$ diverges since we always have $n^2/(n^2-1)>1$.
Therefore, we re-arrange the elliptic integral of third kind in (\ref{eq:ellipticJint})
such that this divergence vanishes \cite{MilneThomson1972}:
\begin{eqnarray}\label{eq:Elliptic3div}
&\Pi_{L}\left(\chi,k,\frac{n^2}{n^2-1}\right)=F_{L}(\chi,k)-\Pi_{L}\left(\chi,k,\frac{k(n^2-1)}{n^2}\right)\\
&+\frac{1}{2}\sqrt{\frac{n^2(n^2-1)}{n^2(1-k)+k}}\ln\left(\frac{\cos\chi\sqrt{1-k\sin^2\chi}+\sqrt{\frac{n^2(1-k)+k}{n^2(n^2-1)}}\sin\chi}{\left|\cos\chi\sqrt{1-k\sin^2\chi}-\sqrt{\frac{n^2(1-k)+k}{n^2(n^2-1)}}\sin\chi\right|}\right) \, . \nonumber
\end{eqnarray}

\section{Solving differential equations using Jacobi's elliptic functions}
Using a suitable coordinate transformation every equation of the form (\ref{eq:geor})
or (\ref{eq:geotheta}) can be transformed into the Legendre form
\begin{eqnarray}\label{eq:Legendrediff}
\left(\frac{\mathrm{d}\chi}{\mathrm{d}\lambda}\right)^2=a(1-k\sin^2\chi),
\end{eqnarray}
where for the time being $\lambda$ shall be an arbitrary real parameter and the
constant $a$ always depends on the coefficient of the highest order. Since we
only look for physical solutions we deal with real quantities and thus in our case
this constant turns out to always be real and positive. Therefore, in the following
we restrict our discussion to this scenario.

Let us now assume that we have a physical setting with initial condition
$\chi(\lambda_{i})=\chi_{i}$ and let us denote the sign of the motion by
$i_{\chi_{i}}=
\mathrm{sign}\left(\left.\mathrm{d}\chi/\mathrm{d}\lambda\right|_{\lambda_{i}}\right)$.
In a first step we separate variables and we integrate using the given initial
conditions:
\begin{eqnarray}
\lambda-\lambda_{i}=\int_{\lambda_{i}}^{\lambda}\mathrm{d}\lambda'=i_{\chi_{i}}\int_{\chi_{i}}^{\chi}\frac{\mathrm{d}\chi'}{\sqrt{a(1-k\sin^2\chi')}} \, .
\end{eqnarray}
Now we move all terms containing the parameter or information on the initial
condition to the left-hand side:
\begin{eqnarray}
\tilde{\lambda}(\lambda)=i_{\chi_{i}}\sqrt{a}(\lambda-\lambda_{i})+\int_{0}^{\chi_{i}}\frac{\mathrm{d}\chi'}{\sqrt{1-k\sin^2\chi'}}=\int_{0}^{\chi}\frac{\mathrm{d}\chi'}{\sqrt{1-k\sin^2\chi'}} \, .
\end{eqnarray}
Using the relation between amplitude $\chi$ and $\tilde{\lambda}$,
$\chi=\mathrm{am} (\tilde{\lambda})$, we see that (\ref{eq:Legendrediff}) is
solved by the Jacobian elliptic $\mathrm{sn}$ function:
\begin{eqnarray}
\sin\chi(\lambda)=\mathrm{sn}(\tilde{\lambda}(\lambda),k)=\mathrm{sn}\left(i_{\chi_{i}}\sqrt{a}(\lambda-\lambda_{i})+\lambda_{\chi_{i},k},k\right),
\end{eqnarray}
where we defined a new quantity $\lambda_{\chi_{i},k}$ that depends on the initial condition.
Analogously for an appropriate coordinate transformation containing $\cos\chi$ we
have $\cos\chi(\lambda)=\cos\mathrm{am} (\tilde{\lambda}(\lambda)) =\mathrm{cn} (\tilde{\lambda}(\lambda),k)$ and thus in
these cases we can write the solution of the general equation in terms of
\begin{eqnarray}
\cos\chi(\lambda)=\mathrm{cn}(\tilde{\lambda}(\lambda),k)=\mathrm{cn}\left(i_{\chi_{i}}\sqrt{a}(\lambda-\lambda_{i})+\lambda_{\chi_{i},k},k\right) \, .
\end{eqnarray}
\section*{References}
\bibliographystyle{iopart-num}
\bibliography{CLensing}
\end{document}